\documentclass[lettersize,journal]{IEEEtran}
\usepackage{amsmath,amsfonts}
\usepackage{algorithmic}
\usepackage{algorithm}
\usepackage{array}
\usepackage[caption=false,font=normalsize,labelfont=sf,textfont=sf]{subfig}
\usepackage{textcomp}
\usepackage{stfloats}
\usepackage[hyphens]{url}  
\usepackage{hyperref}
\hypersetup{breaklinks=true}
\usepackage{verbatim}
\usepackage{graphicx}
\usepackage{cite}
\usepackage[numbers]{natbib}
\usepackage{multirow}
\usepackage{mdframed}
\usepackage{pifont}
\usepackage{enumitem}
\usepackage{booktabs}
\usepackage{xcolor,colortbl}
\usepackage{hyperref}

\definecolor{lightgrey}{rgb}{0.93,0.93,0.93}
\definecolor{darkgreen}{rgb}{0,0.75,0}
\begin{document}

\title{Beyond Functional Correctness: Exploring Hallucinations in LLM-Generated Code}

\author{
\IEEEauthorblockN{
Fang Liu\IEEEauthorrefmark{2},
Yang Liu\IEEEauthorrefmark{2},
Lin Shi\IEEEauthorrefmark{3},
Zhen Yang\IEEEauthorrefmark{4},
Li Zhang\IEEEauthorrefmark{2}*\thanks{* Corresponding author: Li Zhang},
Xiaoli Lian\IEEEauthorrefmark{2},
Zhongqi Li\IEEEauthorrefmark{5},
Yuchi Ma\IEEEauthorrefmark{5}} \\
\IEEEauthorblockA{\IEEEauthorrefmark{2}State Key Laboratory of Complex \& Critical Software Environment (SKLCCSE) \\ School of Computer Science and Engineering, Beihang University, Beijing, China} \\
\IEEEauthorblockA{\IEEEauthorrefmark{3}School of Software, Beihang University, Beijing, China} \\
\IEEEauthorblockA{\IEEEauthorrefmark{4}School of Computer Science and Technology, Shandong University, Qingdao, China} \\
\IEEEauthorblockA{\IEEEauthorrefmark{5}Huawei Cloud Computing Technologies Co., Ltd, China} \\
\IEEEauthorblockA{\{fangliu, liuyang26, shilin\}@buaa.edu.cn, zhenyang@sdu.edu.cn, \{lily, lianxiaoli\}@buaa.edu.cn\\ \{lizhongqi7, mayuchi1\}@huawei.com}}

\maketitle

\begin{abstract}
The rise of Large Language Models (LLMs) has significantly advanced various applications on software engineering tasks, particularly in code generation. Despite the promising performance, LLMs are prone to generate hallucinations, which means LLMs might produce outputs that deviate from users' intent, exhibit internal inconsistencies, or misaligned with the real-world knowledge, making the deployment of LLMs potentially risky in a wide range of applications. Existing work mainly focuses on investigating the hallucination in the domain of Natural Language Generation (NLG), leaving a gap in comprehensively understanding the types, causes, and impacts of hallucinations in the context of code generation. To bridge the gap, we conducted a thematic analysis of the LLM-generated code to summarize and categorize the hallucinations, as well as their causes and impacts. Our study established a comprehensive taxonomy of code hallucinations, encompassing 3 primary categories and 12 specific categories. Furthermore, we systematically analyzed the distribution of hallucinations, exploring variations among different LLMs and benchmarks. Moreover, we perform an in-depth analysis on the causes and impacts of various hallucinations, aiming to provide valuable insights into hallucination mitigation. Finally, to enhance the correctness and reliability of LLM-generated code in a lightweight manner, we explore training-free hallucination mitigation approaches by prompt enhancing techniques. We believe our findings will shed light on future research about code hallucination evaluation and mitigation, ultimately paving the way for building more effective and reliable code LLMs in the future. The replication package is available at \url{https://github.com/Lorien1128/code_hallucination}.
\end{abstract}


\section{Introduction}
Code generation is the process of automatically generating source code based on provided specifications or requirements, which enables developers to save time by reducing manual coding efforts and allows them to focus on higher-level tasks and problem-solving. Moreover, it can aid in ensuring consistency and reducing the risk of human error during development \cite{rasnayaka2024empirical, tabarsi2025llms}. Automatic code generation has been a longstanding challenge in both software engineering and artificial intelligence communities. 
The recent advancements in Large Language Models (LLMs) have significantly propelled this field forward \cite{chen2021evaluatingCodex,roziere2023codellama,chatgpt}. For example, OpenAI's Codex \cite{chen2021evaluatingCodex}, released in 2021, has achieved a success rate of 28.8\% in solving a set of 164 hand-written programming problems. Microsoft's Copilot, a code generation tool powered by Codex, has captured the interest of over 1 million professional developers \cite{euronews}. Moreover, it has shown the potential to speed up coding tasks by up to 55\% \cite{Kalliamvakou}. Subsequently, various code LLMs emerged in both academia and industry, such as Incoder \cite{Fried2023Incoder}, StarCoder \cite{li2023starcoder}, CodeGen \cite{Nijkamp2023codegen,nijkamp2023codegen2}, Code Llama \cite{roziere2023codellama}, DeepSeek-Coder \cite{guo2024deepseek}, DeepSeek-R1 \cite{guo2025deepseek}, ChatGPT \cite{chatgpt}, \textit{etc}. These models have shown impressive performance in many code intelligence tasks, especially in code generation. 

Despite the remarkable success of LLMs, their predictive modeling nature makes them susceptible to generate results that are not correct and reliable \cite{zhong2024can, sharma2024llms}.
They are prone to generate \textit{hallucinations} across various tasks \cite{ji2023survey}. Hallucinations refers to the phenomenon that LLMs might produce nonsensical or unfaithful outputs to the provided source content \cite{ji2023survey,huang2023survey}, which can poses a potential risk in deploying LLMs across various applications. Most existing work mainly focuses on investigating the hallucination for natural language generation (NLG) tasks, for instance, generative question answering \cite{li2021addressing}, abstractive summarization \cite{maynez2020faithfulness}, dialogue generation \cite{huang2020challenges}, \textit{etc}. However, there is still a lack of clarity regarding the specific types of content that LLMs tend to hallucinate during code generation, as well as their causes and potential consequences they may have.
In the integration of the generated code into their development, developers may have the following concerns:
What types of hallucinations does a code LLM typically produce? What are the possible causes and impact of different hallucinations? How are these hallucinations distributed across widely-used LLMs and popular benchmarks?

We argue that such hallucinations also occur in the domain of LLM-based code generation, primary manifesting in two main categories: \textit{knowledge hallucination} and \textit{faithfulness hallucination}, similar to the hallucination categories found in NLG domain \cite{huang2023survey}. 
Given the differences between natural language generation and code generation—where code possesses unambiguous semantics and must adhere to stricter constraints due to its structured nature and specific requirements—\textbf{{we define code hallucination as a phenomenon where an explicit semantic conflict between the generated code and established facts due to the failure of the LLM to retain, recall, or process information accurately from users' requirements, contextual code, or real-world knowledge}}. Consequently, compared to NLG, there may be variations in specific hallucination categories, their causes and impacts in code generation.
The occurrence of code hallucinations may undermine the correctness, performance, maintenance, and even security of the developed software. As a result, the widespread adoption of Code LLMs for code recommendation has the inherent potential to compromise the overall quality and reliability of software. Hence, it is imperative to thoroughly investigate hallucinations in LLM-powered code generation. This will allow us to gain insights into the specific weaknesses that the LLM model may generate, helps us identify code snippets or patterns that are likely to be incorrect or unreliable, providing valuable feedback for improving LLMs. {To clarify terminology, a \textit{code} (or a solution program) refers to a complete, functional program that solves the given problem. A \textit{code snippet} refers to a self-contained segment of code that may constitute either a partial or complete solution program. In this paper, it generally denotes a part (or the entirety) of the LLM-generated code.}

{Driven by the critical gaps in understanding hallucinations in LLM-generated code, we propose three research questions focused on code hallucination. First, while prior studies have explored hallucinations in natural language generation, the unique characteristics of code (\textit{e.g.}, syntax constraints, functional correctness) necessitate a specialized investigation into the categories and distribution of code hallucinations. This motivates \textbf{RQ1 (Distribution of Code Hallucinations)}, which systematically categorizes and quantifies hallucinations across diverse LLMs and benchmarks to establish a foundational understanding of their prevalence. Furthermore, through this research question, we aim to reveal whether certain hallucination categories are model-specific or task-dependent, providing a foundation for targeted improvements. Moreover, \textbf{RQ2 (Hallucination Cause Analysis)} is motivated by the practical imperative to identify factors contributing to code hallucinations. While prior work attributes NLG hallucinations to issues like training data biases or model overconfidence, code generation introduces unique triggers such as ambiguous/incomplete requirements, or knowledge gaps in programming APIs. Investigating these causes helps developers and researchers prioritize mitigation strategies. Finally, understanding the downstream consequences of code hallucinations is crucial for risk assessment and quality review when using LLMs for code generation. Unlike the focus in NLG hallucination, which concentrates on the impact of text consistency, \textbf{RQ3 (Impact of Hallucination)} evaluates the broader impacts of code hallucinations on software quality, including functional correctness, efficiency, and maintainability.}
{These research questions form a holistic framework to characterize code hallucinations, uncover their root causes, assess their implications, and inform future solutions. The interplay between these questions ensures a comprehensive analysis: RQ1 identifies ``what'' code hallucinations exist, RQ2 explains ``why'' they occur, and RQ3 evaluates ``how'' they affect real-world code quality.}

To answer these questions, we conducted a thematic analysis \cite{Cruzes2011thematic} of the LLM-generated code to summarize and categorize the hallucinations presented in it based on the unique characteristics of the code, and also analyze the causes and possible impacts of the code hallucinations.
{Specifically, we collected 3,120 code samples generated by different LLMs on two widely-used code generation benchmarks}, and employ the open coding process to analyze these samples. Finally, we establish a comprehensive taxonomy of hallucinations, which comprises 3 primary categories (\textit{Requirement Conflicting} hallucinations, \textit{Code Inconsistency} hallucinations, and \textit{Knowledge} hallucinations) and 12 specific types of hallucinations, and also summarized their causes and impacts.
Then we conducted a comprehensive investigation and various statistical analyses of these hallucinations from diverse perspectives to gain a deeper understanding of their distributions, causes, and impacts, aiming to explore the prevailing challenges and opportunities in the field of code generation with LLMs. 
The analysis reveals that \ding{182} code LLMs are frequently influenced by a diverse range of hallucinations, with \textit{Requirement Conflicting} being the most prevalent hallucination across all studied LLMs and benchmarks. \ding{183} Both the model's inherent capabilities and the prompt quality contribute to the hallucination. \ding{184} Regarding the impact, incorrect functionality is the most common and direct consequence of a majority of hallucinations, while code readability and efficiency are also typically affected. Therefore, it is imperative to develop effective techniques to mitigate hallucinations during code generation.

Based on our findings, we performed a preliminary experiment to explore training-free hallucination mitigation approaches by prompt enhancing techniques. The results suggest that \ding{185} prompting the LLM to refine requirements or perform chain-of-thought reasoning helps mitigate \textit{Requirement Conflicting} hallucinations, while supplementing prompts with relevant domain knowledge reduces hallucinations stemming from knowledge conflicting hallucinations.

In summary, this paper makes the following contributions:
\begin{itemize}[leftmargin=*]
    \item \textbf{Hallucination taxonomy of LLM-generated code}: We conducted a comprehensive study to analyze the types of content LLMs may tend to hallucinate in code generation, and established a taxonomy of code hallucination categories.
    \item \textbf{Thorough analysis of code hallucinations}: We systematically analyzed the distribution of hallucinations across various LLMs and code generation tasks, and also investigated the causes and impact of code hallucinations.
    \item {\textbf{Hallucination mitigation exploration:} We evaluated three widely-used prompt enhancing strategies to assess their effectiveness in reducing LLM hallucinations, complementing this analysis with in-depth qualitative case studies to uncover the underlying mechanisms.}
    \item \textbf{Implications for addressing code hallucinations}: We deliberated on the findings and implications associated with evaluating and mitigating hallucinations in LLM-generated code.
\end{itemize}

\section{Preliminary}

\subsection{{Terminology}}

To facilitate our definition of hallucination in the context of code generation, we first clarify several key concepts.

\noindent\textbf{Definition 1: Requirement.} The requirement refers to the natural language description about the programming task, as provided in the user's input (prompt) to the LLM.

\noindent\textbf{Definition 2: Contextual Code.} The contextual code encompasses both the code included in the user's input (prompt) and prior code generated by the LLM.

\noindent\textbf{Definition 3: Real-world Knowledge.} The real-world knowledge refers to publicly available and verifiable information derived from human experience or natural activities.

\noindent\textbf{Definition 4: Established Fact.} The established fact refers to objective, clear, verifiable, and context-bounded units of information derived from: (1) requirements (\textit{e.g.}, user-specified input/output format constraints), (2) contextual code (\textit{e.g.}, variable names, API usage given in the contextual code), and (3) real-world knowledge (\textit{e.g.}, mathematical laws).

\subsection{Definition of Code Hallucination}
\label{subsec:definition}

When generating code, LLMs may involve contextual information from various sources, including user requirements, contextual code, and real-world knowledge. For various reasons, the generated code may not always remain consistent with the contextual information and instead exhibit semantic conflicts. We argue that these semantic conflicts can be divided into two main types: (a) conflicts arising from failure to retain or recall contextual information, and (b) conflicts arising from errors in logical reasoning when converting retained information into actual code. In order to maintain consistency with the core meaning of hallucination in the NLG domain, we regard the first type of conflict as \textit{code hallucination}. Given the black-box nature of LLMs, we typically can only analyze the final output (\textit{i.e.}, the generated code), making it often difficult to precisely attribute a conflict to one type or the other. Therefore, for the feasibility of practical classification, we further restrict code hallucination to ``explicit semantic conflict'' between generated code and established facts, where the conflict is immediate and obvious, not dependent on complex logical reasoning. In such cases, the probability that the conflict originates from the second type is small, allowing us to reasonably and clearly classify it as the first type.

Based on the above analysis, we formally define code hallucination as \textbf{a phenomenon where an explicit semantic conflict between the generated code and established facts due to the failure of the LLM to retain, recall, or process information accurately from users' requirements, contextual code, or real-world knowledge}. This discrepancy could arise due to flawed training data sources \cite{lee2022deduplicating, onoe2022entity}, suboptimal choices in training and decoding \cite{zhang2024language, holtzman2020curious}, \textit{etc.}, ultimately producing code that exhibits semantic conflicts with established facts.
Such misalignment manifests through explicit semantic contradictions, serving as a traceable indicator of hallucination.

\begin{figure}[t]
    \centering
    \setlength{\abovecaptionskip}{0.1cm}
    \includegraphics[width=0.85\linewidth]{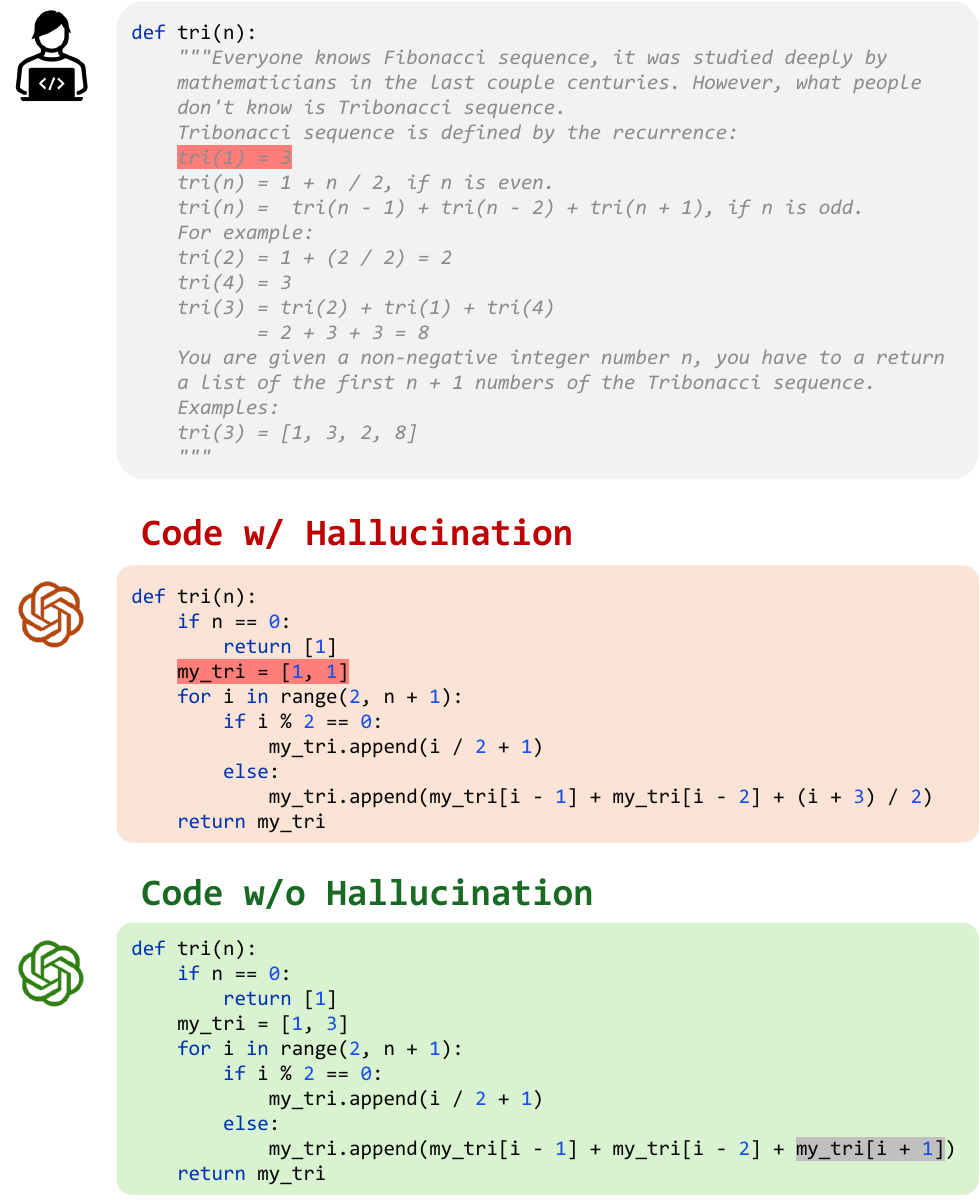}
    \caption{Illustration of code with or without hallucination.}
    \label{fig:hallu_non_hallu}
    \vspace{-0.5cm}
\end{figure}

Specifically, code hallucination can be categorized into two main types: \textit{Faithfulness Hallucinations} and \textit{Knowledge Hallucinations}, analogous to the faithfulness and factuality hallucinations in NLG \cite{ji2023survey,huang2023survey}. \textit{Knowledge Hallucinations} primarily highlights the discrepancies between the generated code and real-world knowledge, and \textit{Faithfulness Hallucination} involves two situations where the generated code conflicts with either the task requirement (\textit{Requirement Conflicting}) or its contextual code (\textit{Code Inconsistency}), {both of which, together with \textit{Knowledge Hallucination}, form the three fundamental types of hallucinations in our taxonomy. Simultaneously, these three types also correspond to the input-conflicting hallucination, context-conflicting hallucination, and fact-conflicting hallucination in NLG \cite{zhang2023siren}.} Given the differences between natural language generation and code generation, there may exist variations in specific hallucination categories, causes, and impacts.

Under this definition, we do not consider indirect (implicit) conflicts that require certain logical reasoning or calculations to manifest as hallucinations, as they are inherent to errors in inference or deduction,
rather than failures to retain, recall, or process established facts accurately. In contrast, code hallucinations reflect a direct misalignment between the model's output and the information that is provided or expected to know.
For example, as illustrated in Figure \ref{fig:hallu_non_hallu},
The requirement of the programming task is to calculate the ``Tribinacci'' sequence according to the defined recurrence rule. Neither of the two samples in the figure fulfill the the requirement correctly. However, only the first code exhibits hallucination. Specifically, the first code incorrectly initializes \texttt{tri(1)} to 1, which directly conflicts with the requirement for \texttt{tri(1) = 3} in the task description. Thus, it belongs to a hallucination. In contrast, the second code fails due to incorrectly accesses out-of-range indexes, or alternatively, obfuscating the logical order of the sequence generation, mistakenly assuming that the subsequent item has already been produced.  
Nonetheless, it is semantically consistent with the requirement, so we do not consider it as an hallucination. In summary, code hallucination focuses on direct semantic conflict without considering naive syntactic errors or logical reasoning/planning errors. 
The detailed differences between them will be explained in Section \ref{subsec:difference}.

\subsection{Difference between Code Hallucinations and Errors}
\label{subsec:difference}

In this section, we highlight the difference between code errors and hallucinations in code generation.
Code with hallucinations does not always mean that the code contains errors. {For instance, if the code's semantic does not conflict with the functional requirements, such as when it includes useless statements, it is unlikely to result in actual errors in most cases.}
In addition, semantic conflicts do not necessarily result in conflicts in execution results.
On the other hand, not all code errors relate to hallucinations. As mentioned in \ref{subsec:definition}, code hallucination focuses on direct semantic conflicts. Therefore, implicit conflicts that require logical deduction to manifest are not within the scope of code hallucination. {Furthermore, if the error is not even caused by semantic conflicts, such as a simple syntactic error, it will certainly not be classified as a hallucination}. To illustrate this more intuitively, below we summarized several typical code error patterns that do not qualify as hallucinations, and we provide an example for each pattern in the online Appendix\footnote{\url{https://github.com/Lorien1128/code_hallucination/blob/master/appendix.pdf}}, which is available in our replication package.

\begin{itemize}[leftmargin=*]
    \item \textbf{Syntactic Error:} Errors that are not related to the semantics of the code. These errors can usually be detected during the compilation phase.
    \item \textbf{Logical Reasoning Error:} LLMs might interpret the requirements in a way that leads to incorrect logic. This can result in logical deductions or calculations in the generated code that, while not directly conflicting with the requirements, are still incorrect.
    \item \textbf{Incomplete Implementation:} This error occurs when {the generated code partially fulfills requirements without directly conflicting with their semantics. For example, a problem might have multiple specific requirements, but the code addresses only a subset of them.}
    \item \textbf{Invalid Generation:} {The generated code is structurally invalid due to syntactic incompleteness or lack of coherent semantics. Unlike \textit{Incomplete Implementation} (where code is structurally sound but fails to meet all functional requirements), \textit{Invalid Generation} involves formally incomplete code, such as missing critical syntax elements (\textit{e.g.}, half-formed function definitions or absent control structures).}
\end{itemize}

\section{Related Work}
\subsection{LLMs for Code Generation}
In recent years, a significant number of LLMs for code-related tasks, especially for code generation, have been proposed \cite{chen2021evaluatingCodex,chatgpt,li2023starcoder,roziere2023codellama}. Codex \cite{chen2021evaluatingCodex} is the earlier representative work to use large generative pre-trained models with up to 12 billion parameters to generate code. It enabled Copilot to deliver real-time coding suggestions, revolutionizing the coding experience. 
The success of Codex has captured the interest of both academia and industry groups in this particular field. As a consequence, various models have emerged. DeepMind proposed AlphaCode \cite{li2022AlphaCode}, which is trained for generating code in real-world programming competitions. Meta proposed InCoder \cite{Fried2023Incoder} and Code Llama \cite{roziere2023codellama},  
Amazon provided CodeWhisperer \cite{CodeWhisperer}, BigCode project proposed StarCoder \cite{li2023starcoder}, DeepSeek-AI introduced DeepSeek-Coder series, OpenAI introduced GPT and ChatGPT series.
The emergence of these models yields remarkable enhancements in the effectiveness of code generation.
To assess the capabilities of LLMs in code generation tasks, researchers have developed various benchmarks.
Notable examples of these benchmarks include HumanEval \cite{chen2021evaluatingCodex}, DS-1000 \cite{lai2023ds1000}, MBPP \cite{austin2021MBPP}, APPS \cite{hendrycks2021apps}, CoderEval \cite{yu2024codereval}, DevEval \cite{DevEval}, \textit{etc}. These benchmarks typically consist of multiple test cases, and are usually accompanied by a few instances to aid in understanding the task description.

\subsection{Hallucination in LLMs}
The term ``hallucination'' has been widely used within the NLG community to describe the generation of text that is nonsensical or deviates from the original source content \cite{ji2023survey,huang2023survey}. 
{In the field of NLP, hallucinations are primarily categorized into two types: \textit{intrinsic hallucinations}, where the generated content contradicts the source content, and \textit{extrinsic hallucinations}, where the generated content cannot be verified from the source input \cite{ji2023survey}. \citet{huang2023survey} further redefined the taxonomy into two main groups: \textit{factuality hallucination} and \textit{faithful hallucination}, aiming to offer a more tailored framework for LLM applications. Factuality hallucination emphasizes the discrepancy between generated content and verifiable real-world facts, typically manifesting as factual inconsistency or fabrication. On the other hand, faithfulness hallucination refers to the divergence of generated content from users' requirements, as well as self-consistency within generated content.
Considering the versatility of LLMs, \citet{zhang2023siren} further refined the definition by categorizing hallucination within the context of LLMs as follows: (1) Input-conflicting hallucination, occurring when the generated content deviates from the original source input; (2) Context-conflicting hallucination, where the generated content contradicts previously generated information; (3) Fact-conflicting hallucination, arising when LLMs produce content that lacks fidelity to established real-world knowledge.
In the context of LLM-powered code generation, the hallucination taxonomy proposed by \citet{zhang2023siren} naturally aligns with the fundamental components of programming tasks: task requirements, contextual information, and real-world knowledge. Building upon this framework, we adapt their taxonomy with code-specific refinements to better characterize and categorize code hallucinations.}

\subsection{Code Hallucination}
{Recent research has increasingly focused on hallucination phenomena in code generation tasks.
\citet{zhang2024llm} studied the phenomena, mechanism, and mitigation of LLM hallucinations in repository-level generation scenario, identifying key hallucination categories such as Task Requirement Conflicts, Factual Knowledge Conflicts, and Project Context Conflicts. In a similar vein, \citet{spracklen2024we} focused on a very specific type of hallucination: package hallucinations, where the generated code erroneously references non-existent libraries, thereby highlighting how model settings, such as training data recency and decoding strategies, can influence these errors.
Other studies have also approached the phenomenon from various angles. For example, \citet{agarwal2024codemirage} defined hallucinated code as code that exhibits one or more defects (\textit{e.g.}, syntactic or logical errors, dead code, security vulnerabilities) and proposes a comprehensive taxonomy of these defects. In contrast, \citet{jiang2024collu} introduced a benchmark that predicts hallucination occurrences by identifying the divergence between LLM outputs and canonical solutions. \citet{rahman2024code} further distinguished hallucinated output from merely incorrect output by emphasizing that hallucinations often include elements that are completely or partially irrelevant to the provided context. Furthermore, \citet{tian2024codehalu} categorized code hallucinations into types such as mapping, naming, resource, and logic hallucinations based on execution-based verification, while \citet{mesbahreducing} underscores the importance of contextual analysis to improve the accuracy of AI-driven code generation.}

{Our research advances previous work in several key aspects. 
On the one hand, we explicitly distinguish code hallucinations from general code errors by emphasizing direct and traceable semantic conflicts with established facts, and addressing the ambiguity in prior studies that overgeneralize \cite{agarwal2024codemirage} or under-specify \cite{rahman2024code} the concept. 
On the other hand, our taxonomy categorizes hallucinations based on their semantic conflict sources (requirements, contextual code, and real-world knowledge), offering finer granularity and more comprehensive scope than repository-level \cite{zhang2024llm} or package-specific \cite{spracklen2024we} classifications. 
Furthermore, we conduct a systematic investigation into how hallucinations manifest across different LLMs and code generation tasks, analyzing their prevalence, patterns, underlying causes, and practical implications. Building on these insights, we explore and discuss several feasible prompt-enhancing strategies for hallucination mitigation.
In summary, our study advances the understanding of hallucination phenomena in code generation by providing a comprehensive analysis that bridges key gaps in current research. Through systematic investigation of previously underexplored dimensions, we offer new insights that both complement and extend the existing body of knowledge in this field.}

\subsection{Evaluation of LLM-Generated Code}
{With the emergence of code LLMs, many researchers began to focus on the generation capabilities of code LLMs. 
Driven by this, a wide variety of benchmarks for evaluating LLM-generated code have been proposed, including standalone function benchmarks \cite{khan2023xcodeeval, chen2021evaluatingCodex, austin2021MBPP, hendrycks2021apps, jain2024livecodebench, chendycodeeval, chen2024ppm}, multi-dependency task benchmarks \cite{zhuo2024bigcodebench, lai2023ds1000}, repository-level task benchmarks \cite{zhang2023repocoder, DevEval, yu2024codereval}, competition-level task benchmarks \cite{hendrycks2021apps, jain2024livecodebench}, and dynamically updated benchmarks \cite{chendycodeeval, chen2024ppm, jain2024livecodebench}, among others.
Based on these benchmarks, numerous studies have examined the quality of the LLM-generated code from different aspects, including security, usability, and especially correctness \cite{jesse2023stupidbugs,liu2023no,yeticstiren2023evaluatingquality,vaithilingam2022expectation,tambon2024bugs}.} 
\citet{jesse2023stupidbugs} explored the extent to which code LLMs are inclined to produce simple, stupid bugs \cite{karampatsis2020often}.
\citet{nguyen2022copilotempirical} assessed the correctness and comprehensibility of GitHub Copilot’s code suggestion. \citet{tambon2024bugs} analyzed the bug patterns in LLM-generated code and their prevalence. \citet{liu2023refining} further conducted an empirical study of ChatGPT-generated code to evaluate its quality and reliability, which also includes an exploration of ChatGPT's self-debugging capability. 
Similarly, \citet{liu2023no} examined the code generated by ChatGPT, with a specific focus on three aspects: correctness, understandability, and security. \citet{dou2024s} analyzed the bug types in code generated by LLMs. 
In contrast to existing research, we conducted the first comprehensive analysis from the perspective of hallucinations to examine the deviations inherent in the LLM-generated code, and also analyzed the code quality issues that can arise from these hallucinations, encompassing most of the quality issues identified in current research \cite{liu2023refining,tambon2024bugs}.

\section{Hallucinations in LLM-Generated Code}\label{empirical_study}

\subsection{Hallucination Taxonomy Construction}

To construct the code hallucination taxonomy, we first collected code samples generated by widely-used LLMs, then adopted the open coding process to analyze these samples, and finally established a taxonomy of code hallucinations.

\subsubsection{Data Collection}
\label{sec:data_collection}

We collect code samples generated by four LLMs on two code generation benchmarks.

\noindent\textbf{Benchmark Selection.} We use two widely-used code generation benchmarks, HumanEval \cite{chen2021evaluatingCodex} and CoderEval \cite{yu2024codereval}, including two popular programming languages and covering common scenarios in current code generation: standalone function and repository-level function generation.

\begin{itemize}[leftmargin=*]
    \item \textbf{HumanEval} consists of 164 standalone hand-written Python coding tasks, covering mathematics, algorithms, reasoning, \textit{etc}. Each task includes a function signature, docstring, ground truth function body as reference, and several unit tests.
    \item \textbf{CoderEval} consists of 230 functions from 43 Python projects and 230 methods from 10 Java projects. These projects are all selected from high star open source projects of various domains, featuring realistic development scenarios. Each task contains the original docstring, the signature, the solution code as reference, and several unit tests to assess the functional correctness of the generated code.  
\end{itemize}

\noindent\textbf{LLM Selection.}
We select four well-known LLMs that have demonstrated great performance in code generation, including CodeLlama \cite{roziere2023codellama}, DeepSeekCoder \cite{guo2024deepseek}, GPT-4 \cite{chatgpt} and DeepSeek-R1 \cite{guo2025deepseek}. This selection encompasses both open-source and closed-source models, as well as general-purpose and code-specialized models. Notably, we include DeepSeek-R1 for its enhanced reasoning capabilities. The diverse characteristics of these models ensure that our evaluation comprehensively represents the current state of advanced LLMs in code generation.

\begin{itemize}[leftmargin=*]
    \item \textbf{CodeLlama} \cite{roziere2023codellama} is an open-sourced code LLM based on the LLaMA \cite{touvron2023llama} architecture, specifically designed for code comprehension and generation tasks. We employ CodeLlama-7B\footnote{\url{https://huggingface.co/codellama/CodeLlama-7b-Instruct-hf}} to generate code for each problem.
    \item \textbf{GPT-4} \cite{chatgpt} is a large-scale, multimodal model developed by OpenAI, exhibiting human-level performance on various tasks, including code generation.
    We employ GPT-4 preview version (gpt-4-0125-preview version\footnote{\url{https://platform.openai.com/docs/models/gpt-4-turbo-and-gpt-4}}) in our evaluation. The specific number of parameters for this version is not publicly disclosed.
    \item \textbf{DeepSeek-Coder} \cite{guo2024deepseek} is a series of code language models trained from scratch, pretrained on 87\% of code and 13\% of natural language. We adopt 1.3B\footnote{\url{https://huggingface.co/deepseek-ai/deepseek-coder-1.3b-instruct}} and 7B\footnote{\url{https://huggingface.co/deepseek-ai/deepseek-coder-7b-instruct-v1.5}} versions of DeepSeek-Coder to generate code.
    \item {\textbf{DeepSeek-R1} \cite{guo2025deepseek} is an open-source reasoning LLM developed by the Chinese startup DeepSeek. It employs advanced reinforcement learning techniques to enhance reasoning capabilities, achieving performance comparable to OpenAI's o1 model across various benchmarks. We used the online API of DeepSeek-R1 (671B)}\footnote{\url{https://api-docs.deepseek.com/zh-cn/guides/reasoning_model}} {for code generation.}
\end{itemize}

\noindent \textbf{LLM Decoding Parameter Settings.}
To minimize the non-deterministic nature of LLMs and reduce variability of our results, for each LLM, we use greedy decoding to generate one code for each problem of the datasets {by setting the \texttt{do\_sample} parameter to \texttt{False}. In this case, the LLM's output is deterministic and remains unaffected by parameters such as \texttt{temperature} and \texttt{top\_p}. Except for the parameters mentioned above, all other parameters of these model use the recommended values in their official documentations.}

\noindent \textbf{Prompt Selection.}
{To avoid introducing additional variables, the prompts fed into the LLMs consisted only of a brief role description and the default problem description from the dataset, with specific details available in the Appendix.}
Then we run these LLMs to generate code for each problem from the above two benchmarks. As illustrated in Table \ref{tab:data_statistics}, we finally collected 3,120 (820+1,150+1,150) code samples generated by these LLMs from HumanEval and CoderEval. Each of the problems in two benchmarks has 5 solutions generated by different LLMs. The last column also presents the average pass@1\footnote{{We calculate the pass@k metric using the method proposed by \citet{chen2021evaluatingCodex}. Specifically, for pass@1, we obtain the result by dividing the number of passed samples by the total number of samples and expressing it as a percentage.}} results for each model across the benchmarks.\footnote{{For the HumanEval dataset, we employ the upgraded EvalPlus\cite{liu2023your} test suite for evaluation.}}

\begin{table}[t]
    \centering \footnotesize
    \setlength{\abovecaptionskip}{0.1cm}
    \caption{Data statistics in manual analysis.}
    \begin{tabular}{l|c|cc|c}
    \toprule
     \rowcolor{lightgrey}
     & \textbf{HumanEval} & \multicolumn{2}{c|}{\textbf{CoderEval}} &   \\
     \rowcolor{lightgrey}
     \multirow{-2}{*}{\textbf{Models}} & \textit{Python} & \textit{Java} & \textit{Python} & \multirow{-2}{*}{\textbf{Avg. pass@1}} \\
    \midrule
    DeepSeek-Coder-1.3B  & 164 & 230  & 230 & 33.33 \\
    DeepSeek-Coder-7B  & 164 & 230  & 230 & 35.42 \\
    CodeLlama-7B  & 164 & 230  & 230 & 26.28 \\
    GPT-4  & 164 & 230  & 230 & 42.47 \\
    DeepSeek-R1 & 164 & 230  & 230 & 48.08 \\
    \midrule
    Total & 820 & 1150 & 1150 & - \\
    \bottomrule
    \end{tabular}
    \label{tab:data_statistics}
    \vspace{-0.3cm}
\end{table}

\subsubsection{Manual Analysis}
\label{sec:manul analysis}
To analyze the hallucination in LLM-generated code, we conducted a thematic analysis \cite{Cruzes2011thematic}. 
According to our previous definition of code hallucination, there are three primary code hallucination categories, \textit{i.e.}, \textit{Requirement Conflicting}, \textit{Code Inconsistency} and \textit{Knowledge hallucinations}, where the first two belongs to the faithfulness hallucinations.
We initially conducted a pilot analysis by randomly sampling 500 code snippets (around 20\% of the total 3,120) to enrich the categories, establish the codebook, and develop an initial taxonomy for code hallucination. 
After obtaining the codebook, the remaining 80\% of the code snippets were labeled by ten annotators to further refine and expand the taxonomy.
In the first phase of labeling (the initial 20\%), both labelers have over four years of experience in Java and Python programming. In the second phase (the remaining 80\%), all the ten labelers possess at least two years of experience in Java or Python programming.

We implemented strict quality control procedures throughout the analysis process. In the initial pilot analysis stage, two experts (co-authors of this paper) with rich Python and Java programming experience and code LLM research experience independently evaluated 500 samples: Given the problem description, reference code provided by the original dataset, execution results (correct or error with detailed error message), they independently review each generated code and identify the existence of hallucinations, including both the specific types as well as the position of the hallucinatory code snippet. {This process takes approximately 60 person-hours.}
{In order to answer RQ2 \& RQ3 and better find ways to deal with code hallucinations, except for the hallucination types, the experts are also asked to identify the possible causes and the impact of the hallucinatory code snippet.} \textbf{It is worth noting that one code may contain multiple hallucinatory code snippets. A single hallucinatory code snippet can have various causes and impact, and annotators are allowed to annotate multiple causes/effects for each hallucinatory code snippet.}
During the labeling process, for the disagreement annotations, the experts and other two of the authors convened for a joint discussion {to incorporate more diverse perspectives and enhance annotation reliability} until a consensus was reached {by an absolute majority. If consensus could not be achieved, the decision was made by the expert with the longest programming experience}. Based on the discussion, we grouped similar categories, resolving conflicts and discrepancies. 
Finally, we organized the annotation results and established the preliminary version of the codebook, illustrating various hallucination types related to LLM-generated code, possible causes and impact of the code hallucinations.

After obtaining the codebook, the remaining 80\% of the samples are labeled independently by 10 participants with Python and Java programming experience, and also familiar with code LLMs, with $\sim $240 person-hours. Each code is annotated by two different participants to eliminate the subjective factors of the annotator. Specifically, they label the samples independently following the same procedure as the previous pilot analysis, aiming to refine and enlarge the taxonomy. If a new hallucination category/cause/impact occurs that the codebook does not cover, the annotator needs to write a description of the hallucination, for further discussions to establish new codes and enhance the codebook and taxonomy. 
We calculated Cohen's Kappa scores pairwise for annotators who independently labeled identical subsets of the dataset across three annotation dimensions: hallucination categories, causes, and impacts. \textit{All Kappa scores ranged from 0.76 to 0.96, exceeding the 0.6 threshold, indicating substantial agreement} \cite{viera2005understanding}.
For the disagreements in annotation, the corresponding participants and a senior author (one of experts in the initial stage, with extensive Python/Java programming experience and code LLM research expertise) convened for a joint discussion until a consensus was reached. Similarly, if consensus could not be reached, the final label was determined by the senior author. More labeling details can be found in Appendix.

\begin{figure*}[t]
    \centering
    \setlength{\abovecaptionskip}{0.3cm}
    \includegraphics[width=0.85\linewidth]{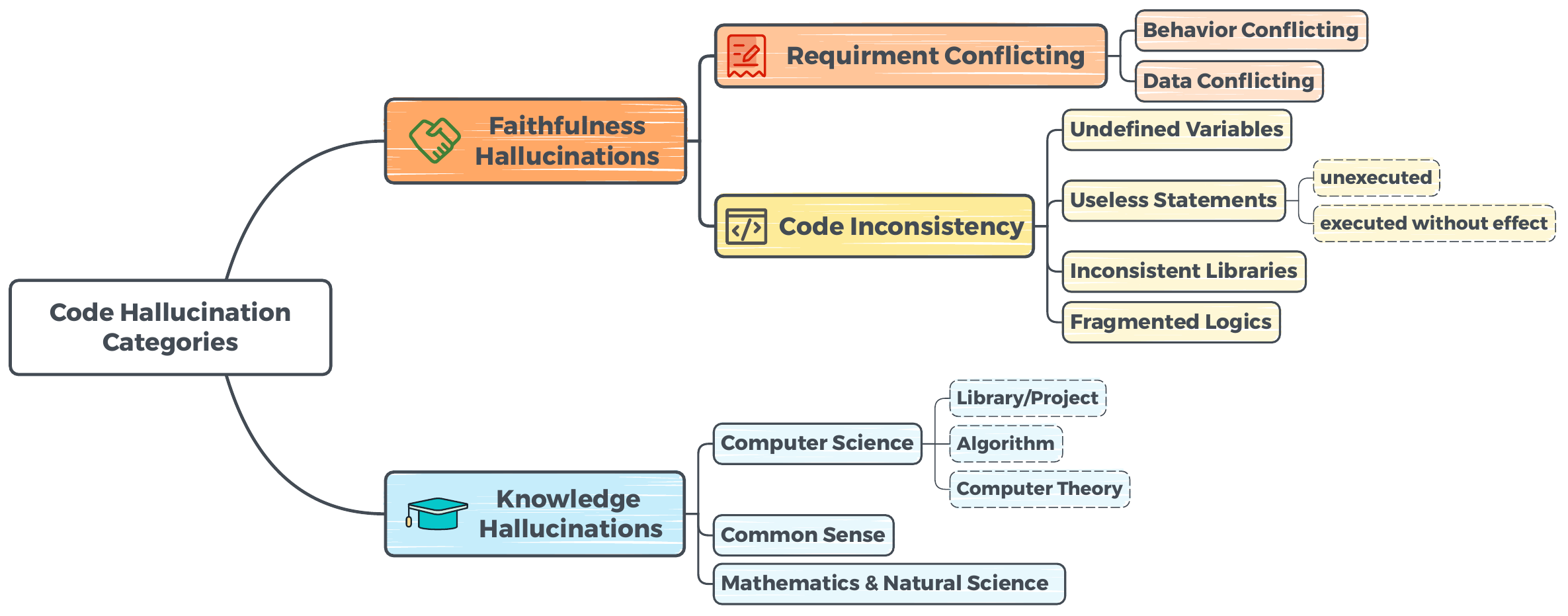}
    \caption{Taxonomy of code hallucinations.}
    \label{fig:hallu_taxonomy}
    \vspace{-0.3cm}
\end{figure*}

Finally, we located 1,212 hallucinatory code snippets (173 from HumanEval, 452 from CoderEval-Python, and 587 from CoderEval-Java) from 1,134 samples out of a total of 3,120 code samples, resulting in a code hallucination taxonomy comprising 3 primary categories and 12 specific categories as leaf nodes, as shown in Figure \ref{fig:hallu_taxonomy}. Detailed explanations of each hallucination category will be provided in Section \ref{sec:hallu_type}. Additionally, the manual study also identified the possible causes and impacts of these hallucinations, which will be discussed in Section \ref{sec:cause} and Section \ref{sec:impact}.

\subsection{Taxonomy of Code Hallucinations}\label{sec:hallu_type}

\begin{figure}[t]
        \centering
        \setlength{\abovecaptionskip}{0.1cm}
        \includegraphics[width=\linewidth]{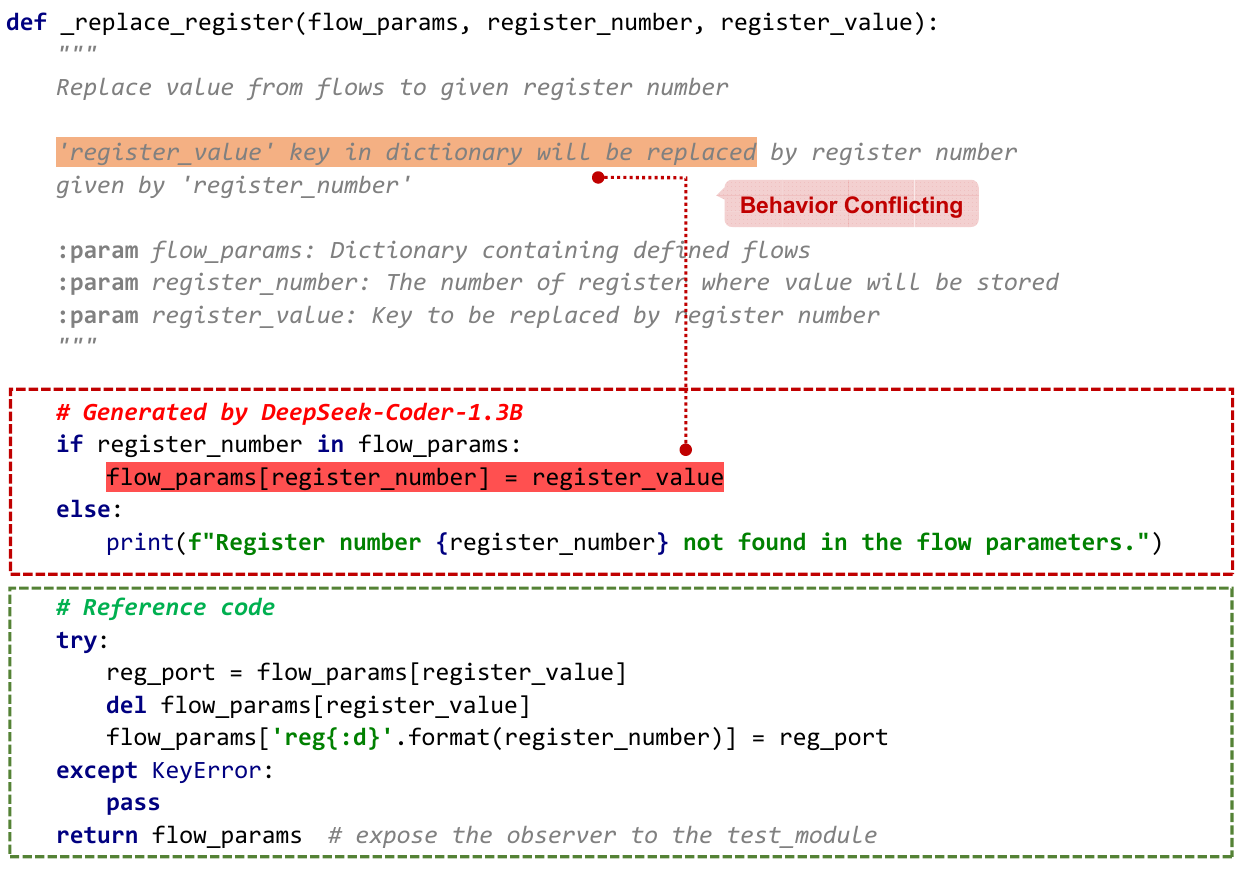}
        \caption{An example of code with \textit{Behavior Conflicting} hallucination.}
        \label{fig:func_req_conflict}
        \vspace{-0.3cm}
\end{figure}

\vspace{1mm}
\noindent \textbf{\textit{I. Requirement Conflicting (39.60\%)}}
\vspace{1mm}

This is the most prevalent and consequential category, referring to the cases where the semantics of the generated code directly conflicts with the given requirements. 
It consists of two subcategories: {\textit{Behavior Conflicting} (35.40\%) and \textit{Data Conflicting} (4.21\%). The primary difference between these two categories lies in the specific content of the requirement that is conflicted.}
{Specifically, \textit{Behavior Conflicting} refers to the cases where the code's functional logic or execution flow deviates from the intended behavior outlined in the requirement (\textit{e.g.}, implementing ``summation'' as ``multiplication''). For example, in Figure \ref{fig:func_req_conflict}, the problem clearly states that \texttt{register\_value} is a \textbf{key} in the dictionary, but the code generated by DeepSeek-Coder-1.3B incorrectly uses \texttt{register\_value} as a \textbf{value} instead, which directly conflicts with the expected behavior specified in the requirements.
This issue may stem from the LLM mistakenly determining the variable's semantic based solely on the meaning of its name.
On the other hand, \textit{Data Conflicting} arises when the data entities in the generated code directly conflicts the requirement-specified values or formats. For example, the requirements might specify a fixed return string (\textit{e.g.}, ``SUCCESS'') under specific conditions, but the generated code instead returns an incorrect value (\textit{e.g.}, ``OK''}).

\begin{figure}[t]
    \centering
    \setlength{\abovecaptionskip}{0.1cm}
    \includegraphics[width=\linewidth]{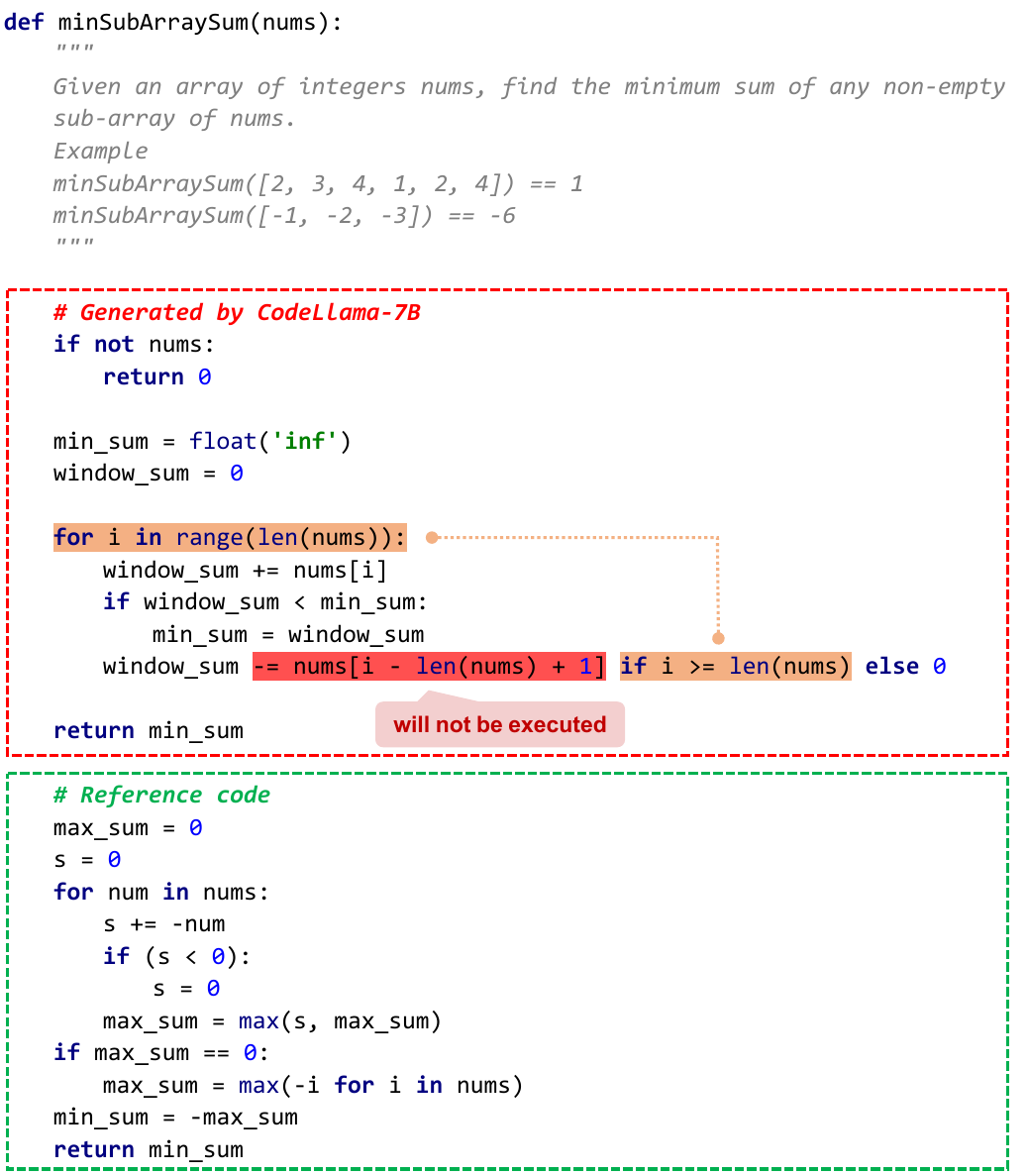}
    \caption{An example of code with \textit{Useless Statements} hallucination.}
    \label{fig:useless_statements}
    \vspace{-0.3cm}
\end{figure}

\vspace{1mm}
\noindent \textbf{\textit{II. Code Inconsistency (25.50\%)}}
\vspace{1mm}

This category refers to situations where direct semantic conflicts arise within the generated code. The \textit{code inconsistency} hallucination typically manifests as \textit{Undefined Variables} (16.91\%), \textit{Useless Statements} (6.60\%), \textit{Inconsistent Libraries} (0.25\%), and \textit{Fragmented Logics} (1.73\%). 
\textit{Undefined Variables} occurs when there is a reference to an undefined variable in the generated code.
\textit{Useless Statements} typically manifests as two scenarios: (1) the generated code contains statements that will not be executed (0.58\%), or (2) the statements will be executed but has no impact on the output (6.02\%).
\textit{Inconsistent libraries} refer to the use of libraries with similar but inconsistent functionality {or the use of different API versions of the same library} in different locations of the code, leading to the failure to achieve the desired objective. For example, \texttt{Matplotlib} and \texttt{Plotly}, as third-party libraries for Python, both have powerful drawing capabilities. LLM may mistakenly mix the two libraries when generating drawing code, using one library in one part of the code and the other in a different part, resulting in semantic conflicts within the code.
\textit{Fragmented Logics} refer to situations where the logical flow of the generated code is disjointed, making it difficult to understand or follow, and increasing the likelihood of errors.
Figure \ref{fig:useless_statements} shows an example of unexecuted hallucination from the \textit{Useless Statements} category. In the code generated by CodeLlama-7B, the loop condition, \texttt{for i in range(len(nums))}, indicates that the maximum value of \texttt{i} is \texttt{len(num) - 1}, but the condition of the \texttt{if} branch in the last line of the loop is \texttt{i >= len(nums)}. This conflict prevents the statement \texttt{-= nums[i - len(nums) + 1]} from being executed.

\begin{figure}[t]
    \centering
    \setlength{\abovecaptionskip}{0.1cm}
    \includegraphics[width=\linewidth]{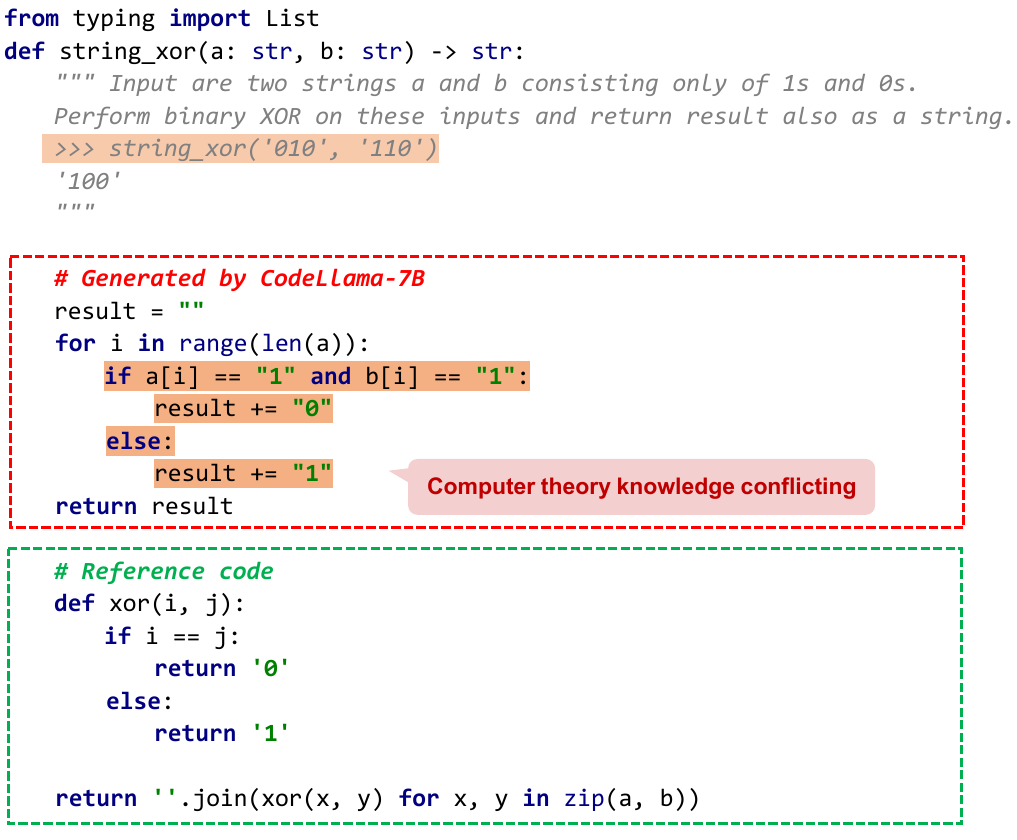}
    \caption{An example of code with \textit{Computer Theory} knowledge conflicting hallucination.}
    \label{fig:computer_conflict}
    \vspace{-0.3cm}
\end{figure}

\vspace{1mm}
\noindent \textbf{\textit{III. Knowledge Hallucinations (34.90\%)}}
\vspace{1mm}

Unlike the previous two categories, \textit{Knowledge Hallucinations} relate to the direct semantic conflict between the generated code and real-world knowledge. Our manual analysis identified three main types of knowledge conflicts: \textit{Computer Science} knowledge conflicting (33.25\%), \textit{Mathematics and Natural Science} knowledge conflicting (1.40\%), and \textit{Common Sense} conflicting (0.25\%).
\textit{Computer Science} knowledge can be further divided into three sub-categories, including \textit{Library/Project} knowledge (25.99\%), \textit{Algorithm} knowledge (4.95\%), and \textit{Computer Theory} knowledge (2.31\%).
Figure \ref{fig:computer_conflict} presents an example of \textit{Computer Theory} knowledge conflicting hallucination, where the model incorrectly assumes that the XOR value of \texttt{a} and \texttt{b} is 1 only when both \texttt{a} and \texttt{b} are 1, which conflicts with the binary calculation rule established in computer theory. {It is worth noting that, although it could also be seen as a logical reasoning error (\textit{i.e.}, the model correctly recalls the XOR definition but errs in translating it into code), this possibility is obviously less likely.}
\textit{Library/Project} knowledge primarily refers to the information contained within the standard library, commonly used third-party libraries, or the specific context of the current project. {Unlike the \textit{Inconsistent Libraries} category, this type of hallucination focuses on single-point conflicts in library/project knowledge, whereas the former concerns multi-point contextual inconsistencies.}
More detailed descriptions of each type of hallucination can be found in the codebook provided in the Appendix of our replication package.

\section{Code Hallucination Analysis}

Based on previous manual analysis results, we further conducted an in-depth analysis to investigate the following research questions.

\begin{figure*}[t]
    \centering
    \setlength{\abovecaptionskip}{0.1cm}
    \includegraphics[width=0.8\linewidth]{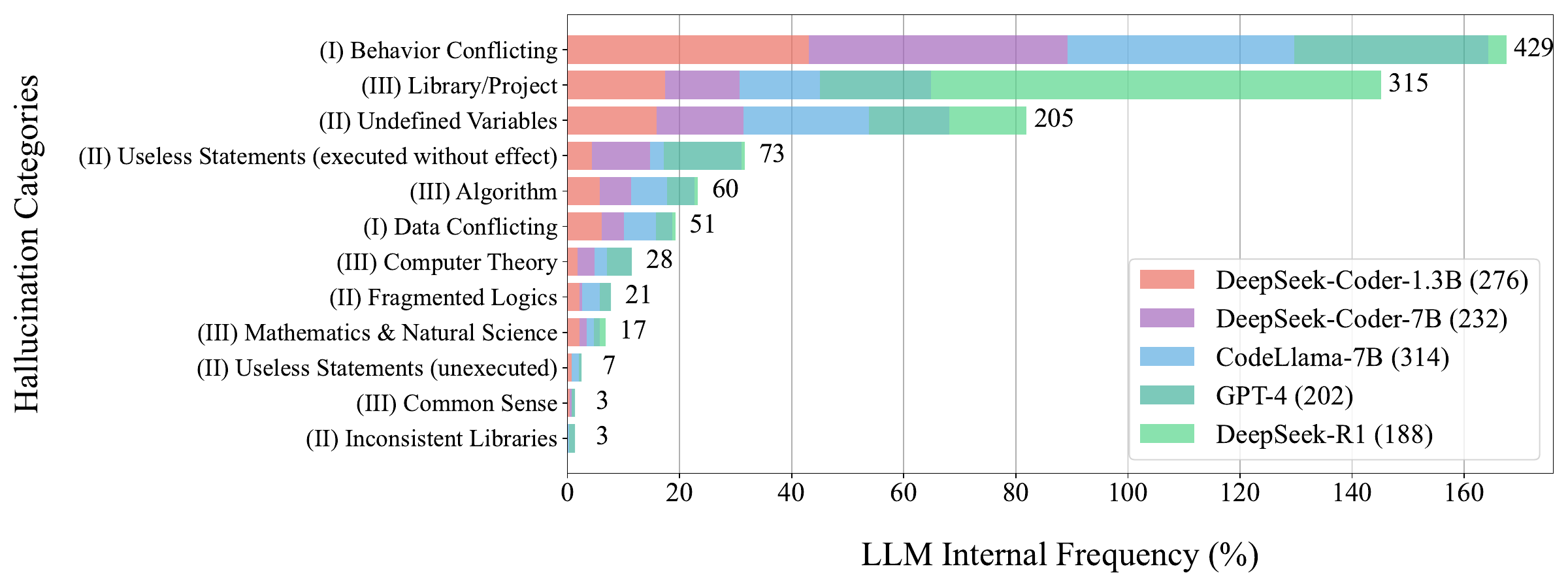}
    \caption{{Overall distribution of code hallucinations.} Frequency denotes the percentage of hallucinations in the specified category relative to the total number of hallucinations produced by a single LLM (denoted by the same color). The number behind each bar represents the total number of this hallucination category, and the number behind each LLM legend represents the total number of hallucinations generated by the model.}
    \label{fig:all_hallu_types}
    \vspace{-0.3cm}
\end{figure*}

\begin{figure*}[t]
    \centering
    \setlength{\abovecaptionskip}{0.1cm}
    \includegraphics[width=\linewidth]{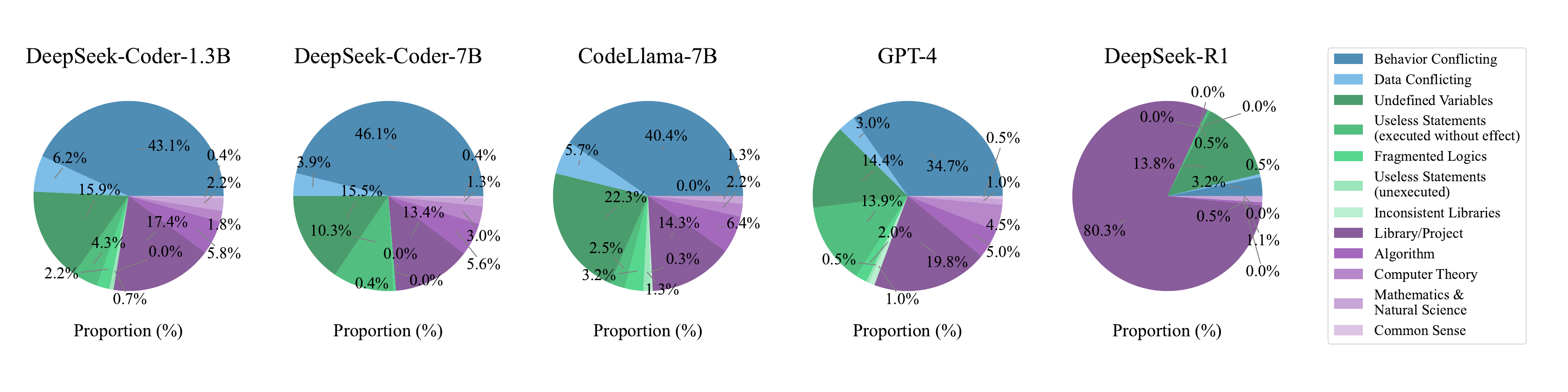}
    \caption{{Distribution of code hallucination in different models.}}
    \label{fig:all_hallu_types_by_model}
    \vspace{-0.3cm}
\end{figure*}

\subsection{RQ1. Distribution of Code Hallucinations.}

In this RQ, we first analyze the overall distribution of various code hallucination categories and their occurrence across different LLMs. We then delve deeper into the hallucination distribution across different benchmarks.

\subsubsection{Overall Distribution of Hallucinations}
Figure \ref{fig:all_hallu_types} and Figure \ref{fig:all_hallu_types_by_model} illustrate the overall distribution of code hallucination categories in different LLMs. 
We can observe that \textit{Behavior Conflicting} is the most frequently occurring type of hallucination across various LLMs, except for DeepSeek-R1. In DeepSeek-R1, \textit{Behavior Conflicting} rank only third among all hallucination categories and are fewer in number than the top two categories. Additionally, \textit{Data Conflicting} is also relatively rare. This suggests that improving reasoning abilities could substantially enhance an LLM's ability to adhere to specified requirements.
Following closely behind are \textit{Library/Project} and \textit{Undefined Variables} hallucinations, both of which are largely identified from CoderEval dataset, as the generated function mostly depends on the contextual information within its own class, project, or other libraries.
Compared with the \textit{Library/Project} knowledge, the hallucination related to the other two subcategories of \textit{Computer Science} knowledge, \textit{i.e.}, \textit{Algorithm} and \textit{Computer Theory}, are relatively less common. 
There are also a certain amount of executed \textit{Useless Statements} hallucination, demonstrating that LLMs struggle to fully capture execution semantics. {However, these issues are less prevalent in DeepSeek-R1, where such hallucinations constitute only a small fraction. This finding implies that increasing model parameters or enhancing reasoning capabilities can effectively alleviate such hallucinations.}

{Although \textit{Inconsistent Libraries} occur the least frequently, they particularity warrants additional attention. As previously defined, this category of hallucination refers to the mixed API use of similar libraries or different versions of the same library. These hallucinations are relatively subtle and difficult to detect, especially for beginners.
Moreover, such inconsistencies can cause runtime errors or prevent the code from achieving its intended functionality. As a result, \textit{Inconsistent Libraries} may lead to higher debugging costs and greater potential risks compared to other categories of hallucinations.}

Regarding the hallucination distribution across LLMs, the current results suggest that the frequency of code hallucinations may be negatively correlated with the model's parameter size. Specifically, the number of hallucinations identified from the code generated by DeepSeek-Coder-1.3B, DeepSeek-Coder-7B, 
and DeepSeek-R1-671B decreases as the parameter size increases. Compared to DeepSeek-Coder-7B, the code generated by CodeLlama-7B exhibit more hallucination issues, even surpassing those found in DeepSeek-Coder-1.3B. This aligns with their performance in code generation, where the pass rate of CodeLlama-7B is lower than both the 1.3B and 7B versions of DeepSeek-Coder, as shown in Table \ref{tab:data_statistics}. The inferior performance of CodeLlama-7B may be attributed to deficiencies in its training data or model design/training strategies, leading to weaker instruction-following capabilities and more frequent hallucinations in generated code compared to a similarly sized model (DeepSeek-Coder-7B). Additionally, the superior performance of DeepSeek-R1 underscores the importance of more advanced LLM training strategies in mitigating code hallucinations. These observations suggest that code hallucinations are not solely influenced by the parameter size of the LLMs, and the underlying mechanisms—including how various factors such as model architecture, training data quality, and parameter size collectively affect hallucination frequency—require further systematic investigation.
Besides, it is interesting to note that, among the five models, DeepSeek-R1, despite exhibiting the fewest hallucinations overall, has the highest proportion of hallucinations in \textit{Library/Project}. This observation further underscores the challenges LLMs face in terms of missing real-world knowledge.
\textbf{Since the number of hallucinations in the last two categories—\textit{Inconsistent Libraries} and \textit{Common Sense}—is quite small, we have omitted them from the following analysis.}

\begin{figure}[t]
    \centering
    \setlength{\abovecaptionskip}{0.1cm}
    \includegraphics[width=\linewidth]{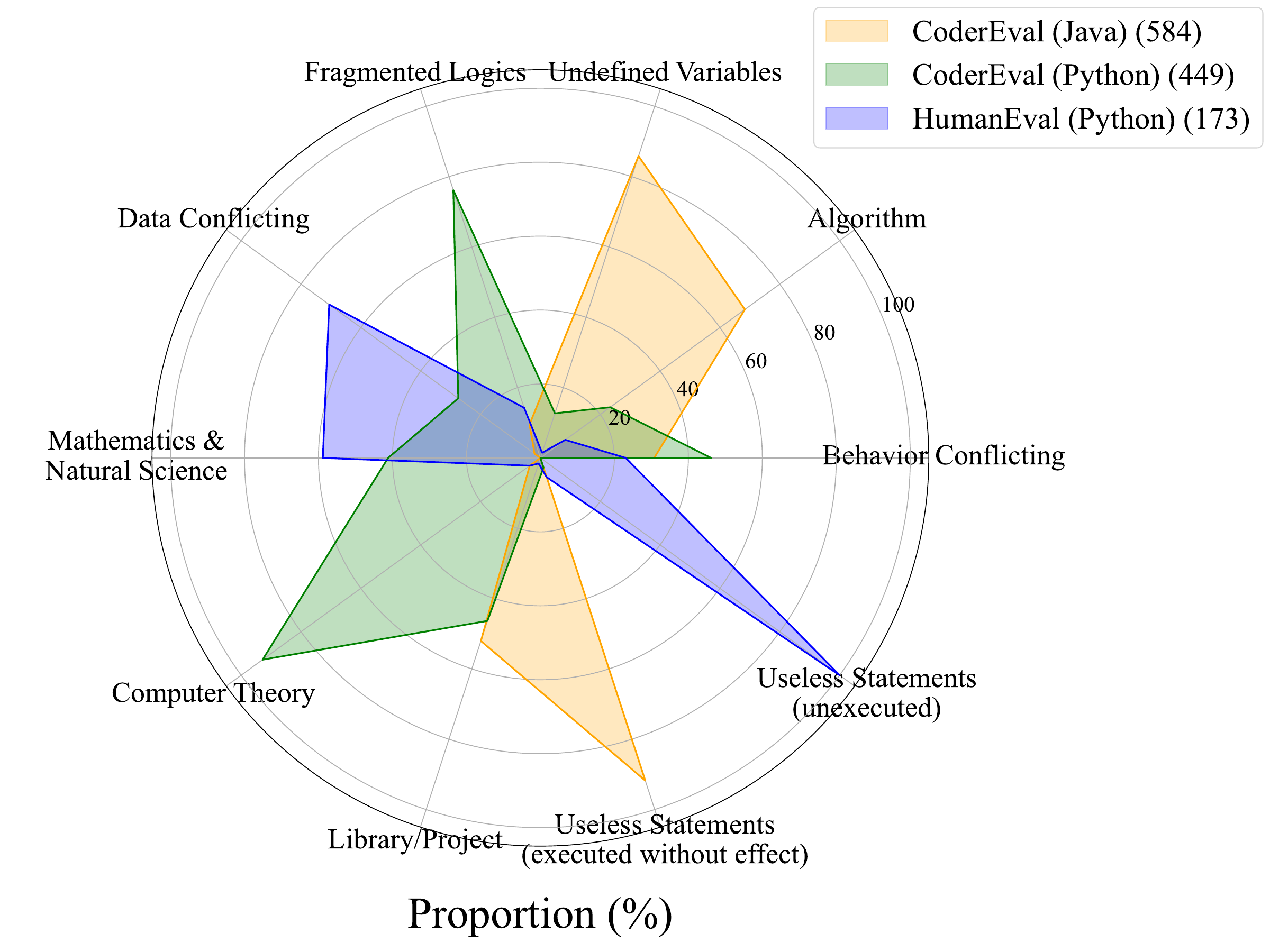}
    \caption{{Distribution of code hallucination on different benchmarks.} Proportion indicate the percentage of hallucinations found in each benchmark relative to the total number of hallucinations in that category. The number behind each legend represents the total number of hallucinatory code snippets within the dataset.}
    \label{fig:dataset_detail}
    \vspace{-0.3cm}
\end{figure}

\begin{figure}[t]
    \centering
    \setlength{\abovecaptionskip}{0.1cm}
    \includegraphics[width=\linewidth]{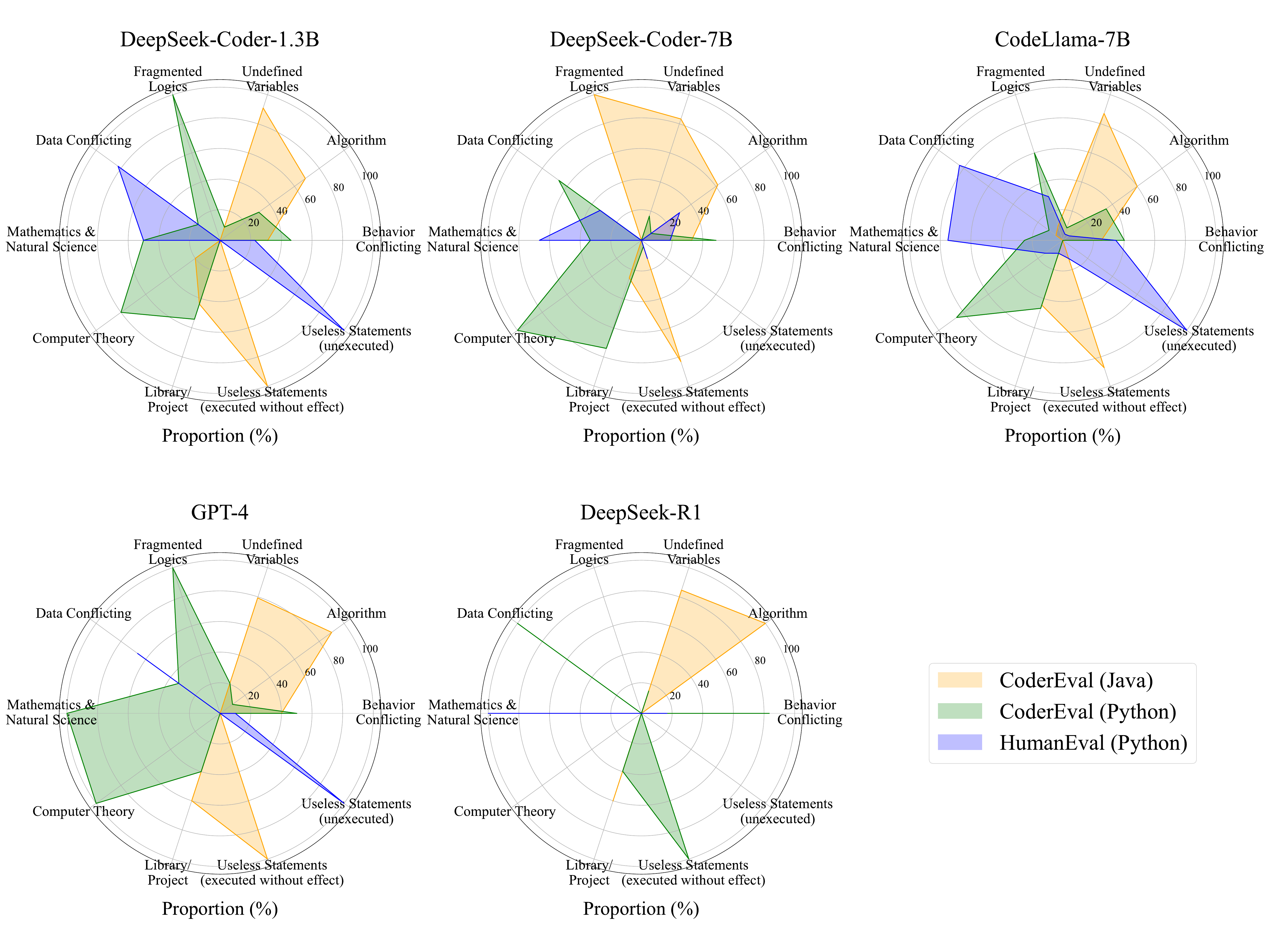}
    \caption{{Distribution of code hallucinations on different benchmarks, divided into separate subplots by model.}}
    \label{fig:dataset_detail_by_model}
    \vspace{-0.3cm}
\end{figure}

\subsubsection{Hallucination Distribution across Benchmarks}
To investigate the hallucination in different as well as real-world code generation scenarios, we employ two representative code generation benchmarks in our manual analysis, covering both standalone function (HumanEval) and repository-level function generation scenarios across Python and Java languages (CoderEval). We present a breakdown distribution of code hallucinations on these two benchmarks in Figure \ref{fig:dataset_detail}. {Specifically, for each benchmark, the results shown in the figure are calculated based on the results of the code generated by all LLMs corresponding to that benchmark.}
We can observe that \textit{Behavior Conflicting} hallucinations commonly occur among all benchmarks.
The programming tasks in HumanEval relate to mathematics and algorithms, which can be solved by standalone functions by accessing only built-in functions and standard libraries in Python. Therefore, most \textit{Mathematics \& Natural Science knowledge conflicting} hallucinations are found in HumanEval dataset. Besides, solving these tasks requires strong logical reasoning ability, LLMs may produce \textit{Useless Statements} when they make incorrect logical deductions, as illustrated in Figure \ref{fig:useless_statements}. Additionally, since LLMs lack a real-time compiler or interpreter during generation, they cannot detect such semantic issues beforehand.  
The programming tasks in CoderEval come from real-world development scenarios, where the solution function could depend on the contextual information within its class, file, project, or the third-party libraries. Therefore, the \textit{Library/Project} hallucinations are mostly found in CoderEval benchmark. Besides, as the prompt only consists of the task requirement, without including the essential contextual information, LLMs tend to generate the \textit{Undefined Variables} or \textit{Useless Statements} as they are unable to determine the correct ones and end up making guesses. This can also lead to \textit{Fragmented Logic} and conflicts with specific \textit{Algorithm} knowledge.
There are also differences in the Python and Java projects in CoderEval benchmark, where more mathematics and computer theory related knowledge is required for solving problems in CoderEval-Python. As a result, LLMs might generate hallucinatory code conflicting the above knowledge in CoderEval-Python.

In addition, we plotted a separate subplot for each model, as shown in Figure \ref{fig:dataset_detail_by_model}. It can be observed that the distribution of code hallucination types across benchmarks remains largely consistent among different models, indicating that our taxonomy is robust and the type distribution is not significantly affected by model variations. The few noticeable discrepancies, such as those between DeepSeek-R1 and the other four models, primarily arise because of the small number of certain hallucination type, making its proportions more sensitive to minor fluctuations in hallucination counts.

We also apply statistical tests on the above results that exhibit obvious differences, and the results demonstrate that the observed differences are significant. Detailed information can be found in the Appendix included in our online replication package. 

\vspace{1mm}
\begin{mdframed}[linecolor=gray,roundcorner=12pt,backgroundcolor=gray!15,linewidth=3pt,innerleftmargin=2pt, leftmargin=0cm,rightmargin=0cm,topline=false,bottomline=false,rightline = false]
  \textbf{Answer to RQ1:} Code LLMs are frequently influenced by a range of hallucinations, with \textit{Behavior Conflicts} being the most prevalent hallucination across all LLMs and benchmarks. {The frequency of hallucinations may be related to various factors of the LLM, including its parameter size—generally speaking, models with larger parameter sizes tend to hallucinate less.} The distribution of hallucination types varies across different code generation scenarios and programming languages. 
\end{mdframed}
\vspace{1mm}

\subsection{RQ2. Hallucination Cause Analysis.}\label{sec:cause}

\begin{figure}[t]
    \centering
    \setlength{\abovecaptionskip}{0cm}
    \includegraphics[width=\linewidth]{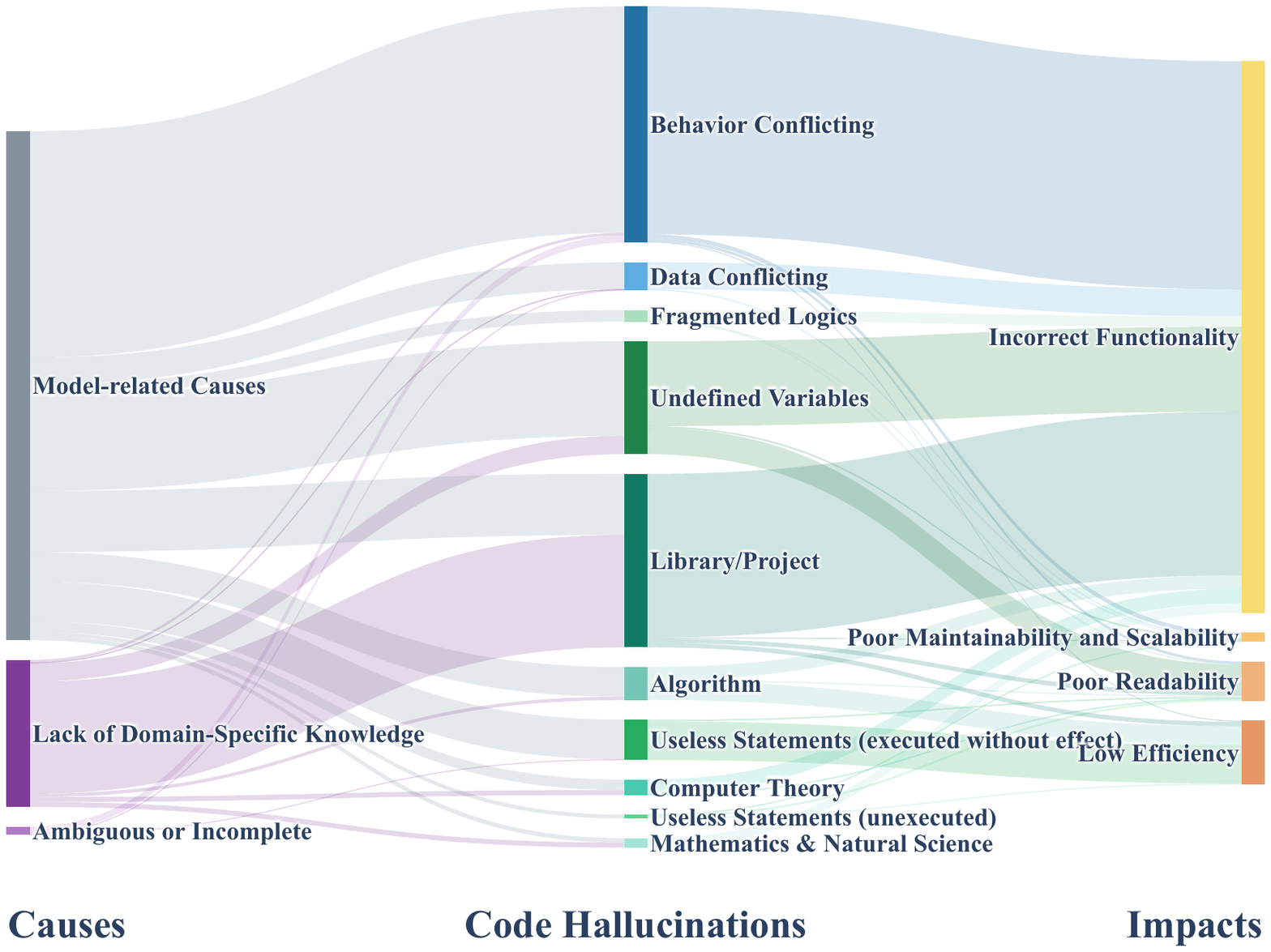}
    \caption{{A visualization of the causes and impacts of various code hallucinations.}\protect\footnotemark[1]}
    \label{fig:cause_type_impact}
    \vspace{-0.3cm}
\end{figure}

Based on our manual analysis results, we have summarized the causes and impacts of hallucinations as shown in Figure \ref{fig:cause_type_impact}.
{Code generation fundamentally involves two components, \textit{i.e.}, the model and the prompt. Therefore, the causes of hallucinations can stem either from model-related or prompt-related factors.}
We categorize the causes of hallucinations into three types: (a) ambiguous or incomplete requirements, (b) lack of domain-specific knowledge necessary for code generation, and (c) model-related factors. The first two causes can be classified as deficiencies in the prompt design, that is, how requirements are articulated to the LLM, representing relatively direct attributions. {Specifically, these two respectively correspond to defects in two sources of established facts, user requirements and real-world knowledge. As for the third source, contextual code, is itself part of the model’s generated output and thus not considered a cause related to the prompt.} Beyond these, other characteristics of prompts, such as text complexity and length, may also indirectly influence the occurrence of hallucinations. To address this, we further conducted a unified modeling analysis of these factors. While diverse model-related factors (\textit{e.g.}, architecture, training data, optimization objectives, \textit{etc.}) result in inherent limitations in the LLMs, given the complexity and impracticality of manually identifying these factors, we categorize them broadly as \textit{Model-related Causes}. Generally speaking, code generation fundamentally involves only two components, the model and the prompt. Therefore, the above cause categories are sufficient to cover the vast majority of cases.

\footnotetext[1]{For the convenience of drawing, for hallucinations containing multiple causes or impacts, we randomly retain one of the causes and impacts. The precise data statistics can be found in our replication package.}

\subsubsection{Distribution of Hallucination Causes}
The detailed information for each cause and its proportion are as follows. {In this RQ, the proportion of each cause category and, in the next RQ, the proportion of each impact category are calculated as the proportion of samples containing hallucinations caused by that cause or resulting in that impact to the total number of samples.} More specific examples can be found in the Appendix.

\begin{itemize}[leftmargin=*]
    \item \textbf{Ambiguous or Incomplete (1.90\%).} Ambiguity or lack of necessary steps or constraints in the requirement. Ambiguity refers to content that can be interpreted in multiple ways. For example, ``merging two integer lists'' could be understood as concatenating the lists or as performing element-wise addition. This can lead to incomplete or incorrect code implementation, as the model may guess the intended behavior, increasing the likelihood of generating misaligned or incorrect code.
    \item \textbf{Lack of Domain-Specific Knowledge (26.24\%).} Code generation often depends on domain-specific knowledge, such as details about classes or functions from a particular codebase. However, this knowledge may have been introduced after the LLM's training period or could pertain to a highly specialized domain not covered in its training data.
    In such cases, the model's parameterized knowledge remains incomplete, resulting in gaps when handling domain-specific code generation tasks. Consequently, if the prompt does not provide the necessary domain-specific knowledge for code generation, the LLM may produce code that conflicts with real-world knowledge. It is important to note that the ``incomplete'' in \textit{Ambiguous or Incomplete} refers to the logical incompleteness of the task description, whereas this type of cause focuses more on the incompleteness of relevant knowledge.
    \item \textbf{Model-related Causes (80.86\%).} The limitations of the LLM itself may contribute to the occurrence of hallucinations, and these limitations can stem from various factors, such as poor-quality training data, a small model parameter size, inherent deficiencies in optimization objective or model architecture \cite{huang2023survey}, \textit{etc}. Given the complexity and impracticality of manually identifying these factors, we collectively classify them as \textit{Model-related Causes}. {In the labeling process, if the prompt shows no obvious flaws, labelers classify the hallucination under this category.}
\end{itemize}

According to the proportion statistics, \textit{Model-related Causes} accounts for the majority of code hallucinations, significantly more than other factors. This suggests that improving the inherent capability of LLMs could be an effective strategy for substantially reducing these hallucinations and enhancing the overall quality of code generation. Specifically, the capability of LLMs can be enhanced through various approaches, such as improving the quality of training data, refining model architecture, optimizing decoding strategies, \textit{etc}. In addition, \textit{Lack of Domain-Specific Knowledge} is also a significant cause of code hallucinations, particularly in code generation for repo-level scenarios.

Furthermore, we also investigate the contributions of these causes to various categories of hallucinations. As shown in Figure \ref{fig:cause_type_impact}, \textit{Model-related Causes} are the primary contributors to almost all categories of hallucinations. In certain hallucination categories (such as \textit{Undefined Variables} and \textit{Library/Project}), factors related to the requirements also play a relatively significant role. {This is especially true for \textit{Library/Project}, as many of the requirements in the CoderEval dataset, based on our manual analysis, do not provide sufficient project-specific knowledge.}
Therefore, it is equally important to write requirements as clearly, accurately, and comprehensively as possible to reduce the occurrence of code hallucinations when using LLMs to generate code.

{To further validate the reliability of our cause annotations, we conducted additional experiments focusing on two prompt-related causes, \textit{Ambiguous or Incomplete} and \textit{Lack of Domain-Specific Knowledge}. Specifically, we selected the Deepseek-Coder-1.3B model for verification. For \textit{Lack of Domain-Specific Knowledge}, we randomly sampled 144 tasks (with 95\% confidence level and 5\% margin of error) from the 230 Python tasks in CoderEval. Among these, 9 hallucinatory samples were identified as partially or fully caused by this cause, and 7 of which involved \textit{Library/Project} hallucinations. After enriching the prompts with relevant repository code retrieved via a RAG-based method, only 2 samples remained exhibiting \textit{Library/Project} hallucinations. Similarly, for \textit{Ambiguous or Incomplete}, we manually optimized all 4 corresponding prompts in HumanEval whose generated code contained hallucinations arising from this cause, and after regeneration, only 1 sample remained hallucinatory. The reduction of hallucinations in the newly generated samples confirm that the identified causes are accurate, further supporting the soundness of our analysis and conclusion in this RQ.
}

\subsubsection{Impact of Prompt Length and Complexity}

{In addition to the three direct causes of hallucinations mentioned above, code hallucinations may also be indirectly influenced by other characteristics of the prompt. Among these, the complexity and length of the prompt are two fundamental features. Prompts that are too complex or too long may exacerbate the LLM's tendency to forget or confuse established facts, thereby increasing the likelihood of code hallucinations.}
Our manual analysis (Section \ref{sec:manul analysis}) also requires labelers to label the text complexity of prompts on a scale of 1-3, with higher values indicating greater complexity \cite{fitzgerald2015important, fitzgerald2016examining}. 
Considering the characteristics of the code generation task specifically, we established four criteria for evaluating complexity: whether the prompt (1) necessitates repeated readings to comprehend its intent, (2) involves multiple reasoning steps, (3) requires specialized domain knowledge or newly introduced definitions, or (4) exhibits syntactical complexity (e.g., nested clauses or prepositional phrases).
Labelers then determined the complexity level based on these criteria. The specific criteria and examples can be found in the codebook in the Appendix.
Regarding the prompt length, we used the \texttt{tiktoken} library to calculate the token length of each prompt.

\begin{figure}[t]
    \centering
    \setlength{\abovecaptionskip}{0.1cm}
    \includegraphics[width=0.9\linewidth]{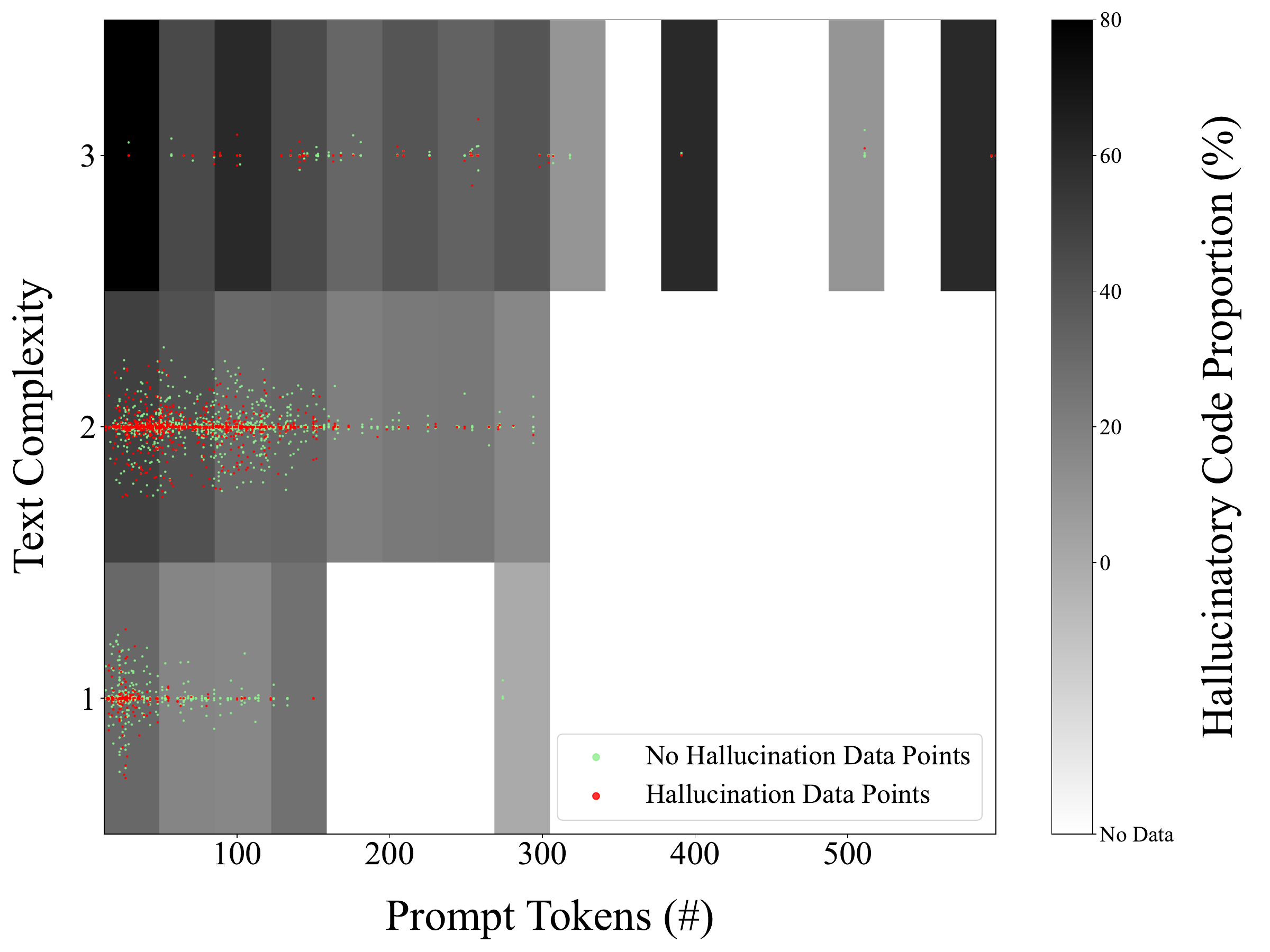}
    \caption{{Heatmap of the proportion of hallucinatory code across different text complexity and length of prompts, along with a scatter plot of the distribution of prompts corresponding to hallucinatory and non-hallucinatory code. The darker the color, the higher the proportion of hallucinatory code, while the white areas represent regions with no data points. To improve visual discrimination of discrete complexity levels (1–3), we slightly dispersed the text complexity values of identical data points. The corresponding figure generated from raw data can be found in the Appendix.}}
    \label{fig:prompt_analyze}
    \vspace{-0.3cm}
\end{figure}

To investigate the potential relationship between prompt complexity and length, we conducted a joint analysis of these factors' influence on code hallucinations using the scatter plot presented in Figure \ref{fig:prompt_analyze}.
{It can be observed that the probability of hallucinatory code occurrence is indeed influenced by both the complexity and length of the prompt. Furthermore, the distribution of the scatter plot indicates a strong positive correlation between length and complexity.} Regarding the impact of prompt length, the probability of hallucinatory code occurrences initially fluctuates and decreases as prompt length increases. This decline may be attributed to the fact that longer prompts allow for more detailed and clarified problem descriptions, which helps LLM better understand the requirement and reduce the likelihood of generating hallucinations.
However, once the length reaches approximately 350, the probability of hallucinations then rises significantly. This may be due to the LLM's difficulty in capturing key points when the prompt becomes too lengthy.
In contrast, the impact of prompt complexity shows a clearer trend: higher complexity levels consistently correlate with increased hallucination rates. This suggests that complex logic poses a greater challenge for LLMs in accurately grasping the intent of the requirements. We performed a p-test on the correlations mentioned above between these three variables to support our qualitative analysis, with detailed results provided in the Appendix.

Additionally, when the complexity is fixed, it appears that changes in prompt length do not significantly affect the proportion of hallucinatory code. This suggests that the effect of prompt length on hallucination rates may be mediated by complexity. To validate this hypothesis, we performed a Bootstrap mediation effect significance test on the original data, resulting in a 95\% Bootstrap confidence interval of $(0.0113, 0.0725)$, with both bounds greater than 0, indicating that the mediation effect is significant. In other words, the impact of prompt length on hallucinatory code is largely explained by its relationship with complexity.

\vspace{1mm}
\begin{mdframed}[linecolor=gray,roundcorner=12pt,backgroundcolor=gray!15,linewidth=3pt,innerleftmargin=2pt, leftmargin=0cm,rightmargin=0cm,topline=false,bottomline=false,rightline = false]
  \textbf{Answer to RQ2:} {The inherent capabilities of LLMs and deficiencies in prompt design are major contributors to code hallucinations. Other prompt-related factors, particularly the complexity of the prompt, also play a crucial role in causing hallucinations.} Enhancing the inherent capabilities of LLMs, such as the reasoning capability, and writing requirements clearly, comprehensively, and concisely could be effective strategies for reducing code hallucinations and improving code quality.
\end{mdframed}
\vspace{1mm}

\subsection{RQ3. Impact of Hallucination.}\label{sec:impact}

As shown in Figure \ref{fig:cause_type_impact}, the impacts of the hallucinations can be summarized into four categories. The detailed information for each impact and its proportion are as follows:

\begin{itemize}[leftmargin=*]
    \item \textbf{Incorrect Functionality (95.30\%).} This is the most obvious and direct impact of hallucination. When the generated code conflicts with the requirement, contextual code or real-world knowledge, the functionality of the code is highly likely to be incorrect. {This could fundamentally undermines the usability and reliability of the software system, as users encounter unexpected behaviors when interacting with the erroneous implementations. From the debugging perspective, such hallucinations introduce debugging significant challenges, requiring developers to cross-validate the code against requirements, contextual information, or domain-specific knowledge to locate the root cause, increasing troubleshooting complexity.}
     \item \textbf{Poor Readability (14.27\%).} Certain hallucinations may come with complex code logic or irrelevant code statements, leading to poor readability. {This may hinder collaborative development, as team members struggle to understand and modify opaque implementations. Furthermore, this also extends debugging timelines and elevates the risk of introducing new defects during maintenance.}
    \item \textbf{{Low Efficiency} (5.20\%).} Some categories of hallucinations may not impact the functionality of the code itself but can introduce redundant logic {or unnecessary variables, potentially reducing code efficiency. While remaining the functionality, these inefficiencies may degrade system responsiveness and resource utilization, which probably become critical in latency-sensitive systems. Identifying and resolving such issues demands performance profiling expertise to distinguish hallucination-induced overheads from legitimate computational costs. This additional complexity not only extends debugging efforts but also increases maintenance burdens. It is important to note that compiler optimizations may mitigate such inefficiencies during execution. However, code execution efficiency depends on numerous interacting factors, making it difficult to isolate their effects experimentally. Furthermore, since our research focuses on the semantics of the generated code rather than the compiled machine code, we do not account for compiler optimizations in our analysis.}
    \item \textbf{Poor Maintainability and Scalability (3.88\%).} {Similarly, hallucinatory code may exhibit poor coding style and structure, potentially violating the principles of high cohesion and low coupling or increasing the complexity of function interfaces. This can negatively impact maintainability and scalability.} {These hallucinations systematically undermines system evolution, making feature extensions error-prone and system upgrades prohibitively expensive. Resolving such issues often requires comprehensive architecture review rather than localized fixes, demanding cross-component impact analysis that dramatically increases maintaining efforts.}
\end{itemize}

According to the proportion statistics and Figure \ref{fig:cause_type_impact}, we can observe that 8 out of 10 hallucination categories result in \textit{Incorrect Functionality}. 
The most frequent hallucination category is \textit{Behavior Conflicting}, where the generated code deviates from the expected behavior in requirements and may produce erroneous outputs, thus making incorrect functionality the most frequent impact.
Besides, a large percentage of samples contain \textit{Knowledge} and \textit{Code Inconsistency} hallucinations, like \textit{Library/Project} knowledge conflicting, \textit{Algorithm} knowledge conflicting, and \textit{Undefined Variables}, may also result in incorrect functionality. 

In addition, {we also present the distribution of general code errors versus hallucinations across all generated samples in Table \ref{tab:error_hallucination},} and further calculate the pass@1 scores for samples with and without hallucination as well as for each individual hallucination category, as shown in Table \ref{tab:passk_statistics}. 
{The results show that code containing hallucinations is more prone to errors. Meanwhile, non-hallucinatory code may also contain errors, but these are unrelated to hallucinations. The findings intuitively illustrate the distinctions between hallucinations and general coding errors from a data-driven perspective, and demonstrate that hallucinatory code—across most hallucination categories—exhibits lower functional correctness than hallucination-free code, thereby supporting our earlier analysis.}
Therefore, mitigating code hallucinations will greatly improve the functional correctness of LLM-generated code. 

\begin{table}[t]
    \centering \footnotesize
    \setlength{\abovecaptionskip}{0.1cm}
    \caption{{2×2 contingency table of code errors and hallucinations. The percentages in parentheses represent the proportion of the total sample.}}
    \begin{tabular}{lccc}
    \toprule
    & \textbf{w/ Error} & \textbf{w/o Error} & \textbf{Total} \\
    \midrule
    \textbf{w/ Hallucination} & 1,063 {\scriptsize(34.07\%)} & 71 {\scriptsize(2.28\%)} & 1,134 {\scriptsize(36.35\%)} \\
    \textbf{w/o Hallucination} & 899 {\scriptsize(28.81\%)} & 1,087 {\scriptsize(34.84\%)} & 1,986 {\scriptsize(63.65\%)} \\
    \textbf{Total} & 1,962 {\scriptsize(62.88\%)} & 1,158 {\scriptsize(37.12\%)} & 3,120 \\
    \bottomrule
    \end{tabular}
    \label{tab:error_hallucination}
    \vspace{-0.4cm}
\end{table}

\begin{table}[t]
    \centering \footnotesize
    \setlength{\abovecaptionskip}{0.1cm}
    \caption{{Pass@1 scores in different categories of hallucinations.}}
    \begin{tabular}{lc}
    \toprule
    \textbf{Category} & \textbf{Pass@1} \\
    \midrule
    \textbf{w/o Hallucination} & \textbf{54.73} \\
    \midrule
    \textbf{w/ Hallucination} & \textbf{6.26} \\
    Computer Theory & 0.00 \\
    Common Sense & 0.00 \\
    Inconsistent Libraries & 0.00 \\
    Mathematics \& Natural Science & 0.00 \\
    Undefined Variables & 0.49 \\
    Data Conflicting & 1.96 \\
    Library/Project & 4.76 \\
    Fragmented Logics & 4.76 \\
    Useless Statements (executed without effect) & 5.48 \\
    Behavior Conflicting & 7.46 \\
    Algorithm & 33.33 \\
    Useless Statements (unexecuted) & 57.14 \\
    \bottomrule
    \end{tabular}
    \label{tab:passk_statistics}
    \vspace{-0.3cm}
\end{table}

Following \textit{Incorrect Functionality}, the impacts of \textit{Poor Readability} and \textit{Low Efficiency} are also common, typically arising from hallucination categories other than \textit{Requirement Conflicting}. For example, \textit{Useless Statements} (executed without effect) may not lead to functional incorrectness, but their unnecessary occurrence can reduce efficiency, and these useless statements also could negatively affect the code's readability. Additionally, code with \textit{Algorithm} knowledge conflicting hallucination may use incorrect or suboptimal algorithms, resulting in high time complexity and, consequently, low execution efficiency. Thus, while the impacts of different hallucinations may appear similar, their underlying mechanisms can differ.

\vspace{1mm}
\begin{mdframed}[linecolor=gray,roundcorner=12pt,backgroundcolor=gray!15,linewidth=3pt,innerleftmargin=2pt, leftmargin=0cm,rightmargin=0cm,topline=false,bottomline=false,rightline = false]
  \textbf{Answer to RQ3:} Incorrect functionality is the most frequent impact, as most code hallucinations can lead to this consequence, suggesting that mitigating code hallucinations will greatly improve the functional correctness of LLM-generated code. Moreover, code hallucinations can negatively affect code readability and execution efficiency, which are also important factors to address for improving overall code quality.
\end{mdframed}
\vspace{1mm}

\begin{figure*}[t]
    \centering
    \setlength{\abovecaptionskip}{0.1cm}
    \includegraphics[width=0.85\linewidth]{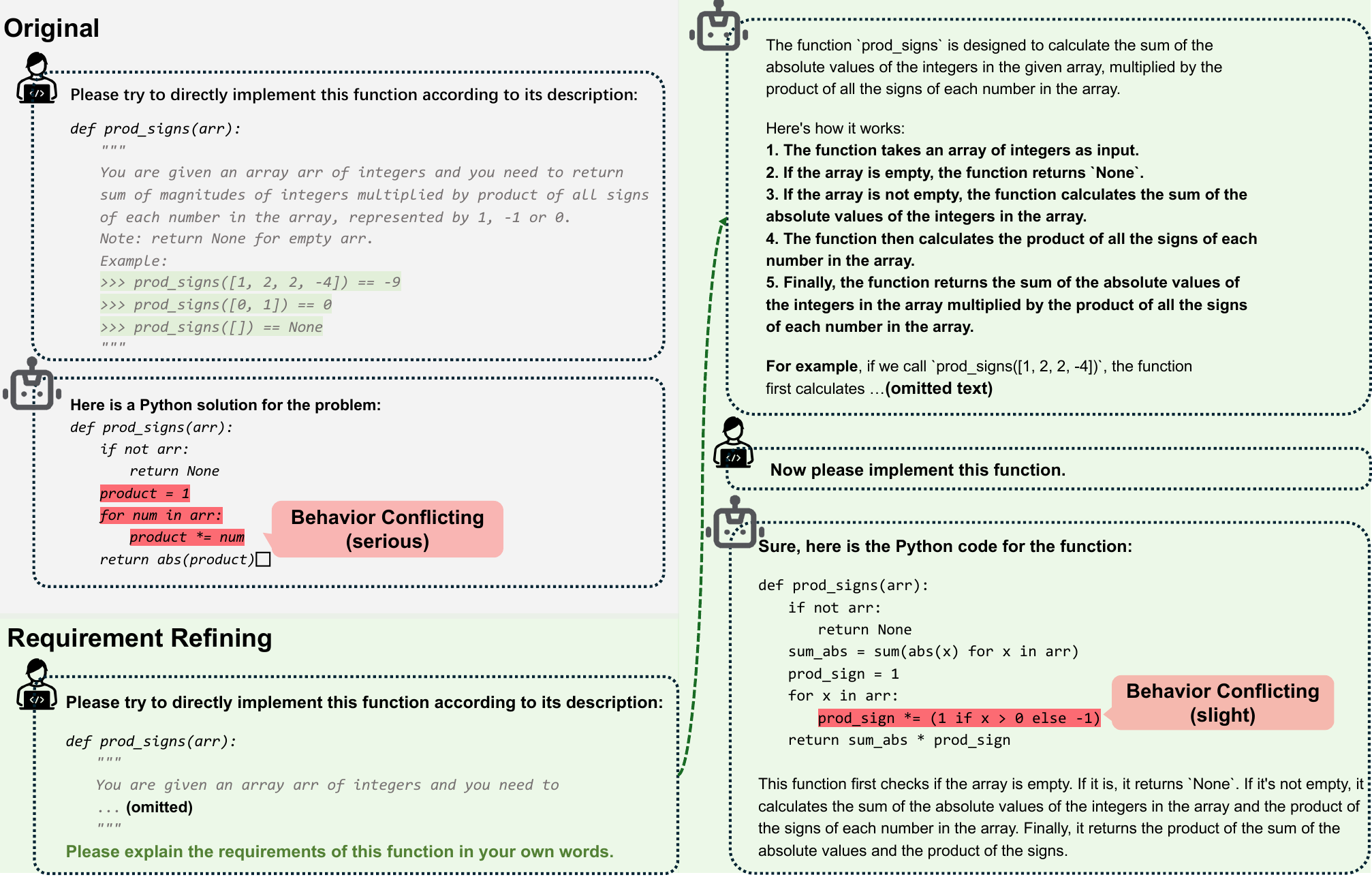}
    \caption{An example of using Self-Refine for hallucination mitigation with DeepSeek-Coder-1.3B.}
    \label{fig:mit_exp_1}
    \vspace{-0.3cm}
\end{figure*}

\section{Discussion}

\subsection{Hallucination Mitigation Exploration}
\label{sec:mitigation_exp}

Previous analysis have demonstrated that code generated by LLMs frequently suffers from various hallucination issues. Addressing these issues is crucial for ensuring the quality and reliability of the generated solutions. Building upon our established taxonomy of hallucinations, along with their corresponding causes and impacts, we argue that it would be effective to design tailored methods to address specific types of hallucinations. Unlike traditional code generation approaches, LLMs have the potential to learn from the prompt and solve problems without resorting to model parameter updates, such as Self-Repair \cite{gupta2020synthesize,zhang2023self}, Self-Refine \cite{madaan2024self-refine,Cycle}, Retrieval-Augmented Generation (RAG) \cite{lewis2020retrieval}, Chain-of-Thought (CoT) \cite{wei2022chain,cov} prompting techniques, \textit{etc}. Compared to the training-based approaches, these methods offer a more flexible and adaptive way for addressing the issues. 
Thus, in this section, we explore and discuss several feasible prompting enhancing strategies for hallucination mitigation in code generation.
For example, \textit{Requirement Conflicting} hallucination, which may arise from misunderstanding requirements or poor reasoning during instruction following, can potentially be mitigated by prompting LLMs to refine requirements or apply chain-of-thought reasoning. For \textit{Code Inconsistency} hallucination, chain-of-thought prompting techniques can also be used to help the LLM create more reasonable plans for implementations. Additionally, post-processing methods can be applied to eliminate common {\textit{code inconsistency} conflicts. For example, static code analysis tools can be used to detect \textit{Undefined Variables} and replace them with the most similar defined variable.} As for the \textit{Knowledge hallucination}, relevant {real-world} knowledge can be retrieved as context and provided to the LLM with Retrieval Augmentation Generation (RAG) techniques, helping to compensate for any gaps in the model's real-world knowledge.

To demonstrate the feasibility of the above ideas and provide references for future research, we experimented with several feasible prompting enhancing strategies for hallucination mitigation, including \textbf{Self-Refine}, \textbf{CoT} and \textbf{RAG}.  
{We chose DeepSeek-Coder-1.3B as the LLM for our experiments, with generation parameters consistent with those in Section \ref{sec:data_collection}. For Self-Refine and CoT, as previously discussed, these strategies have stronger potential to mitigate \textit{Requirement Conflicting} and \textit{Code Inconsistency} hallucinations. Therefore, for these two strategies, we selected the HumanEval dataset for evaluation due to its high prevalence of these specific hallucination types. The prompt formats are the same as in Figure \ref{fig:mit_exp_1} and the first hallucination mitigation example in the Appendix respectively. Similarly, for RAG-based strategy, we employed CoderEval (Python) dataset for evaluation. We used a sliding window approach ({window size: 20 lines, step size: 2 lines}) with the OpenAI \texttt{p50k\_base} tokenizer to retrieve semantically similar code snippets from the repository, and then constructed prompts with a maximum token length of 1,000 in the same format as shown in Figure \ref{fig:mit_exp_2}.}

\begin{table*}[t]
    \centering \footnotesize
    \setlength{\abovecaptionskip}{0.1cm}
    \caption{{Pass@1 scores and manual analysis results for the three prompting enhancing strategies used to mitigate hallucination.}}
    \label{tab:hallucination_mit}
    \resizebox{1.0\textwidth}{!}{ 
    \begin{tabular}{l|l|l|c|ll|cc|cc|cc}
        \toprule
        \rowcolor{lightgrey}
          & & & & \multicolumn{2}{l|}{\textbf{Hallucinatory Samples}} & \multicolumn{2}{c|}{\textbf{\textit{Requirement Conflicting}}} & \multicolumn{2}{c|}{\textbf{\textit{Code Inconsistency}}} & \multicolumn{2}{c}{\textbf{\textit{Knowledge Hallucinations}}} \\
        \rowcolor{lightgrey}
          \multirow{-2}{*}{\textbf{Dataset}} & \multirow{-2}{*}{\textbf{Strategy}} & \multirow{-2}{*}{\textbf{Pass@1}} & \multirow{-2}{*}{\textbf{Samples}} & \textit{quantity} & \textit{proportion} & \textit{quantity} & \textit{proportion} & \textit{quantity} & \textit{proportion} & \textit{quantity} & \textit{proportion} \\
        \midrule
        \multirow{3}{*}{\textbf{HumanEval}} & Origin & 61.59 & \multirow{3}{*}{115} & 30 & 26.09\% & 28 & 84.85\% & 2 & 6.06\% & 3 & 9.09\% \\
        & Self-Refine & 54.27~\scriptsize\textcolor{red}{7.32$\downarrow$} & & 16~\scriptsize\textcolor{darkgreen}{14$\downarrow$} & 13.91\%~\scriptsize\textcolor{darkgreen}{12.18\%$\downarrow$} & \textbf{9} & \textbf{56.25\%} & 3 & 18.75\% & 4 & 25.0\% \\
        & CoT & 51.83~\scriptsize\textcolor{red}{9.76$\downarrow$} & & 18~\scriptsize\textcolor{darkgreen}{12$\downarrow$} & 15.65\%~\scriptsize\textcolor{darkgreen}{10.44\%$\downarrow$} & \textbf{15} & \textbf{83.33\%} & 3 & 16.67\% & 0 & 0.0\% \\
        \midrule
        \multirow{2}{*}{\textbf{CoderEval (Python)}} & Origin & 14.78 & \multirow{2}{*}{144} & 62 & 43.06\% & 38 & 57.58\% & 7 & 10.61\% & 21 & 31.82\% \\ 
        & RAG & 26.52~\scriptsize\textcolor{darkgreen}{11.74$\uparrow$} & & 29~\scriptsize\textcolor{darkgreen}{33$\downarrow$} & 20.14\%~\scriptsize\textcolor{darkgreen}{22.92\%$\downarrow$} & \textbf{17} & \textbf{58.62\%} & \textbf{3} & \textbf{10.34\%} & \textbf{9} & \textbf{31.03\%} \\
        \bottomrule
    \end{tabular}}
\end{table*}

We first execute the code generated under these three strategies and obtained their pass@1 results. Then, we randomly sampled a subset for manual evaluation with a 95\% confidence level and a 5\% confidence interval. Our quality control procedures followed the same rigorous standards established in Section \ref{sec:manul analysis}, and the results are shown in Table \ref{tab:hallucination_mit}. In the table, ``Origin'' refers to baseline code generation using the original prompt template from Section \ref{sec:data_collection}. For the ``Hallucinatory Samples'' columns, ``quantity'' and ``proportion'' refer to the number of hallucinatory samples and their percentage relative to total samples, respectively. And for the last six columns, ``quantity" and ``proportion'' refer to the number of hallucinations of each type and the proportion of that type of hallucination among all hallucination types.
{The results show that these three strategies substantially reduce hallucinations in their respective target datasets, with particularly effective mitigation of \textit{Requirement Conflicting}. For RAG, we observe a notable reduction in \textit{Knowledge Hallucinations} on the CoderEval (Python) dataset, which aligns with our expectations. However, Self-Refine and CoT strategies exhibited decreased pass@1 performance, which is primarily attributed to the limitation to smaller LLMs' (1.3B) constrained reasoning capacity, where multi-step processing amplifies initial errors through error propagation \cite{shim2024cot}. This further explains why CoT did not have a significant impact on the number of \textit{Code Inconsistency} hallucinations.} {We conducted a further investigation into the tasks that led to a decrease in pass@1. For Self-Refine, among the 17 tasks that originally passed but failed after prompt enhancement, all 17 corresponding samples under the original prompt did not contain hallucinations. For CoT, the numbers are 22 and 21, respectively. In summary, the reduction in pass@1 caused by prompt enhancement mainly originates from samples that were originally free of hallucinations.}

To further understand the mechanisms behind the effectiveness of these strategies, we provide several specific examples for qualitative analysis. Figure \ref{fig:mit_exp_1} presents an example of employing Self-Refine for mitigating \textit{Behavior Conflicting}. In this example, the task is to calculate the sum of the absolute values of each number in the given array, and then multiplying it by the product of their signs. The original generated code contains a serious \textit{Behavior Conflicting}, which directly calculates the absolute value of the product of each number, rather than the product of their signs. To mitigate this issue, we prompt the LLM to explain the requirement, enabling it to refine the requirement on its own before generating the code. As seen from the figure, after the refinement, although the code generated still exhibits the \textit{Behavior Conflicting} (only considering 1 or -1 as possible sign values, which conflicts with the expected code behavior specified in the problem description, where the values should be 1, -1, or 0.). However, compared to the original code, the hallucination is notably milder. Additionally, when the input list does not contain a 0, the new generated code is functionally correct.

\begin{figure*}[t]
    \centering
    \setlength{\abovecaptionskip}{0.1cm}
    \includegraphics[width=0.85\linewidth]{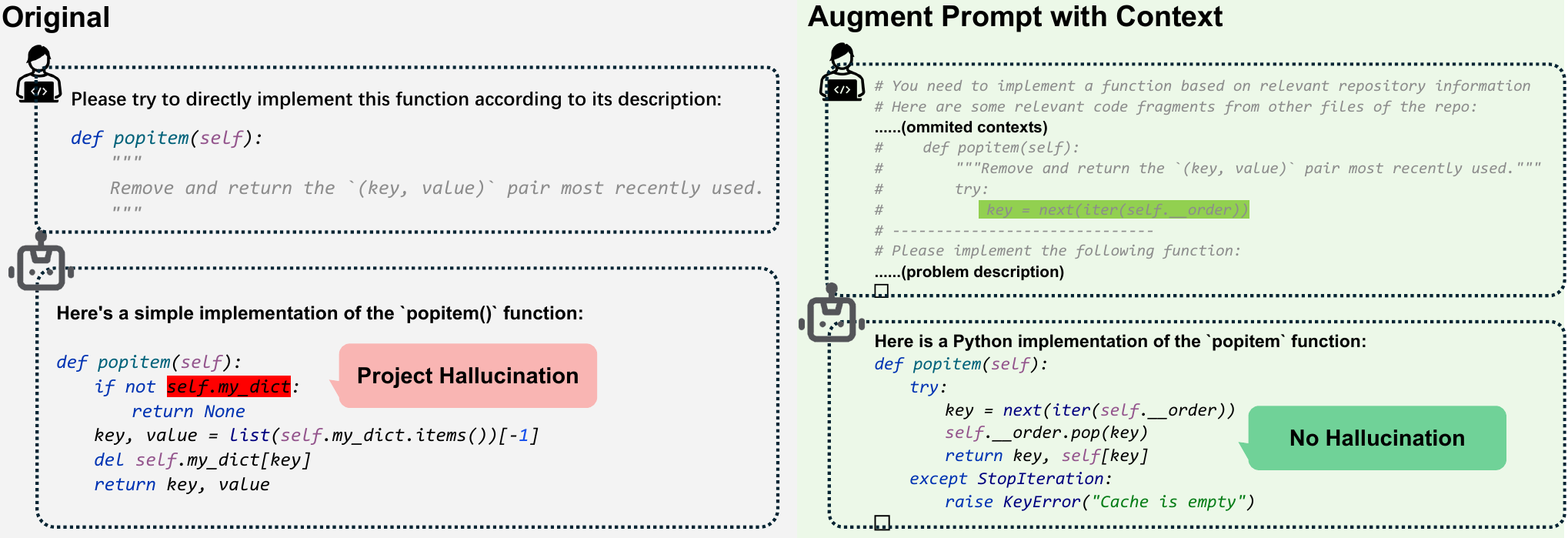}
    \caption{An example of using RAG for hallucination mitigation with DeepSeek-Coder-1.3B.}
    \label{fig:mit_exp_2}
    \vspace{-0.3cm}
\end{figure*}

Figure \ref{fig:mit_exp_2} illustrates an example of using RAG to augment the prompt for mitigating \textit{knowledge hallucination}. Originally, LLM struggles to understand the requirements precisely without the relevant project knowledge. For instance, they might incorrectly use a attribute/method that does not exist in the class (\textit{i.e.}, using \texttt{self.my\_dict} instead of \texttt{self.\_\_order}), resulting in the appearance of \textit{knowledge hallucination}. Actually, the project-specific context can provide essential semantic information for understanding the functionality as well as offering valuable insights into the design and implementation of the target code. It is worth noting that the functionality of \texttt{popitem} method in another class in the context happens to be complementary to the target function, with much of the underlying logic and implementation being largely similar. Therefore, by referring to the context, the LLM can generate code that closely aligns with the project's requirements. More examples can be found in the Appendix included in our online replication package.

Based on the above analysis, we believe that the prompt enhancing techniques present a promising direction for hallucination mitigation. 
Our hallucination taxonomy serves as a useful guidance for developers in selecting optimization methods based on their specific needs and preferences. For instance, if a developer prioritizes time complexity, and as shown in Figure \ref{fig:cause_type_impact}, the low efficiency is largely due to the LLM's lack of algorithmic knowledge, the developer can focus on retrieving relevant algorithms and incorporating them as context when interacting with LLMs.
However, the effectiveness of the prompt enhancing techniques still heavily relies on several factors, including the LLM's inherent ability to understand and following requirements, its logical reasoning skills, and the accurate extraction of relevant real-world knowledge. 
{Therefore, improving the LLM's inherent abilities through advanced training techniques (\textit{e.g.} reinforcement learning \cite{chatgpt-RL}), improved decoding strategies \cite{dhuliawala2023chain, li2023inference}, model architecture optimization \cite{rebuffel2022controlling, li2023batgpt}, and multi-agent interaction \cite{du2023improving}, as well as leveraging external tools for knowledge retrieval and fact-checking \cite{xie2023adaptive, weijia2023replug}, are also worthwhile areas of study for future research.}

\subsection{Implications}
In this section, we discuss the implications of our findings for researchers and developers in software engineering and LLM fields.

\subsubsection{Implications for Researchers}
In our study, we have observed that code LLMs also experience and are frequently influenced by hallucinations. 
While most evaluation metrics and benchmarks in existing code generation research prioritize the functional correctness of the code, incorporating measures and benchmarks to detect hallucinations is crucial for providing a more comprehensive and nuanced assessment of the quality and reliability of the LLM-generated code. 
Considering the enormous workload and potential subjectivity associated with manual hallucination detection, it is necessary to develop automated methods for detection. One promising direction is to leverage advanced LLMs directly for detection. In this approach, the hallucination taxonomy, representative examples for each category, the generated code, and its contextual information can be provided to the LLM, which then determines whether a hallucination exists and identifies its type. To improve accuracy and reliability, more sophisticated mechanisms can be introduced, such as ensemble voting across multiple LLMs, a coarse-to-fine classification strategy, multi-agent debates to determine outcomes, or incorporating a critic model to evaluate the reasonableness of the predicted type. Additionally, based on our manually annotated dataset, the dataset can be further expanded and used to fine-tune an LLM, allowing it to learn code hallucination detection in a supervised manner.

Regarding hallucination mitigation in code generation, according to our findings in RQ2, most of the hallucinations are caused by model-related causes. 
Coupled with our hallucination mitigation exploration study in previous subsection, it is imperative to improve the reasoning capabilities of LLMs by advanced training/fine-tuning techniques.
For example, employing reinforcement learning algorithms during the training can teach the model to think productively using its chain-of-thought, thereby improving its chain-of-thought reasoning ability during inference \cite{chatgpt-RL,roit2023factually,wu2024fine,RLCoder}.
Additionally, integrating RAG techniques and domain-specific knowledge \cite{zhang2023repocoder,zhao2024retrieval,Dataflow-RAG,graphCoder} warrants further investigation, as they hold promise for boosting the LLM's performance by leveraging external information to support and refine its reasoning process. 

Furthermore, in this paper, we primary focus on studying hallucinations in the NL2Code generation task. However, hallucination occurrences and distribution may vary across different code-related tasks, such as code translation \cite{roziere2020unsupervised}, unit test generation \cite{yuan2023no}, program repair \cite{xia2023automated}, code review \cite{CoderReviewer,Llama-reviewer}, \textit{etc}. This presents an opportunity for further research to explore the characteristics and patterns of hallucinations in these tasks. By gaining a deeper understanding of the underlying mechanisms, researchers can develop task-specific strategies and techniques to mitigate hallucinations, improving the accuracy and reliability of code LLMs for specific code-related tasks.

\subsubsection{Implications for Developers}
The emergence of the LLM-as-a-service community has become a prominent trend, allowing developers to easily deploy and utilize various LLMs in their daily development tasks. {To reduce the hallucinations, according to our findings in RQ1, the model with larger parameter size tend to hallucinate less, thus selecting an LLM with robust performance capabilities is crucial}.
Additionally, the quality of prompts plays a significant role. Developers should craft prompts that are clear, concise, and complete, ensuring they provide all necessary context to guide the model effectively. Besides,  applying detailed constraints or rules also help guide the LLM to generate code stays within certain boundaries and reduces the likelihood of hallucinations. 

Finally, rigorous result-checking and iterative prompt refining processes are also essential for validating and improving the generated results.
Given that hallucinations are not always easy to identify and can lead to quality issues, result checking \cite{peng2023check,chern2023factool} becomes even more critical. Besides, the developers should also carefully review and thoroughly test the generated code to ensure its logical correctness and consistency with the given requirements. If any issues are detected during this process, refining the prompts based on the feedback obtained from testing and review can be beneficial in identifying and mitigating hallucinations.

\subsection{Threats to Validity}
\noindent\textbf{Threats to external validity} relate to the generalizability of our code hallucination taxonomy and findings. Our empirical study specifically targets Python and Java programming tasks with varying difficulty and domains from two widely-used code generation benchmarks, covering both standalone function and repository-level function generation scenarios. The LLMs evaluated in this study are advanced, representative, and widely used in software engineering research. They encompass both open-source and closed-source models, as well as general-purpose and code-specialized models, ensuring our conclusions generalize to other LLMs. During manual analysis, we annotated LLM-generated code sequentially by model. Notably, the final three (of four) LLMs introduced no new hallucination types, further validating both the completeness of our taxonomy and its applicability across diverse models. Additionally, the consistent distribution of hallucination proportions observed across all evaluated LLMs underscores the robustness of our findings.
Nonetheless, it would be interesting to explore the hallucinations in other programming languages and code generation scenarios as well.

\noindent\textbf{Threats to internal validity} relate to the subjectivity of our manual analysis. The labeling of a code snippet's hallucination categories, causes, and impact is somewhat subjective, and different annotators may have varying determinations of the same code snippets. To address this, we initially created a codebook with precise guidelines for each hallucination categorization, cause, and impact based on an pilot analysis. Additionally, we organized discussions and meetings to resolve any conflicts or uncertainties that arose. This comprehensive process significantly enhances the reliability and quality of the taxonomy of hallucination categories, causes, and impact in LLM-generated code resulting from our manual analysis. However, despite these efforts, providing rigorous categorization criteria to completely eliminate subjectivity remains unrealistic. Whether in natural or programming languages, semantics are inherently continuous and lack clear-cut discrete divisions. Attempting to impose artificial boundaries would require numerous ad hoc exceptions, which is impractical and could undermine the generality and robustness of the taxonomy. A promising future direction is to leverage LLMs as annotators to automatically assign categories to given samples. Even when a taxonomy lacks strictly quantitative criteria, such an approach may yield more objective, consistent, and reproducible annotations. Nevertheless, due to the limited interpretability of LLMs themselves, the reliability and validity of this method remain important questions for further exploration.

\noindent\textbf{Threats to construct validity} relate to the variability in LLM-generated code in our manual analysis due to decoding strategies and different prompts. To address this issue, regarding the decoding strategy, we use greedy decoding to generate one code for each problem of the datasets, ensuring LLMs produce consistent outputs for the same input, thereby reducing variability. 
As for the prompt design, to ensure a fair evaluation across all models, we employ a consistent prompt structure for all LLMs and tasks within the benchmarks. This structure includes two core components: (1) a role description (\textit{e.g.}, ``I want you to play the role of a programmer, and please ...") to contextualize the LLM's objective, and (2) the programming task description provided in the dataset, which outlines specific programming requirements.
According to our findings in Section \ref{sec:mitigation_exp}, prompt engineering can have significant impact on the quality of the generated code, which is a promising research direction for code hallucination mitigation.

\section{Conclusion}
In this paper, we present an empirical study on code generation hallucinations produced by LLMs. Through a comprehensive manual analysis, we developed a taxonomy of hallucinations through open coding and iterative refinements. 
Further investigation reveals that code LLMs are frequently affected by hallucinations, particularly those that conflict with users' requirements. The code hallucinations are influenced by both the model's inherent capabilities and the quality of the prompts, leading to incorrect functionality, poor readability, poor maintainability, and low efficiency.
We also explore training-free approaches for mitigating hallucinations through prompt enhancement techniques. Finally, we discuss the implications of our study and propose future research opportunities to improve the reliability and trustworthiness of LLMs in code generation. To foster further research, we have made the replication package publicly available at \url{https://github.com/Lorien1128/code_hallucination}.

\section*{Acknowledgement}
This work is supported by the National Natural Science Foundation of China Grants Nos. 62302021, 62577007, 62332001, 62502283, and the State Key Laboratory of Complex \& Critical Software Environment (Grant Nos. CCSE-2025ZX-09 and CCSE-2024ZX-14).

\bibliography{ref}

@article{ji2023survey,
  title={Survey of hallucination in natural language generation},
  author={Ji, Ziwei and Lee, Nayeon and Frieske, Rita and Yu, Tiezheng and Su, Dan and Xu, Yan and Ishii, Etsuko and Bang, Ye Jin and Madotto, Andrea and Fung, Pascale},
  journal={ACM Computing Surveys},
  volume={55},
  number={12},
  pages={1--38},
  year={2023},
  publisher={ACM New York, NY}
}

@article{zhang2023siren,
  title={Siren's Song in the AI Ocean: A Survey on Hallucination in Large Language Models},
  author={Zhang, Yue and Li, Yafu and Cui, Leyang and Cai, Deng and Liu, Lemao and Fu, Tingchen and Huang, Xinting and Zhao, Enbo and Zhang, Yu and Chen, Yulong and others},
  journal={arXiv preprint arXiv:2309.01219},
  year={2023}
}

@article{chen2021evaluatingCodex,
  title={Evaluating large language models trained on code},
  author={Chen, Mark and Tworek, Jerry and Jun, Heewoo and Yuan, Qiming and Pinto, Henrique Ponde de Oliveira and Kaplan, Jared and Edwards, Harri and Burda, Yuri and Joseph, Nicholas and Brockman, Greg and others},
  journal={arXiv preprint arXiv:2107.03374},
  year={2021}
}

@article{li2022AlphaCode,
  title={Competition-level code generation with alphacode},
  author={Li, Yujia and Choi, David and Chung, Junyoung and Kushman, Nate and Schrittwieser, Julian and Leblond, R{\'e}mi and Eccles, Tom and Keeling, James and Gimeno, Felix and Dal Lago, Agustin and others},
  journal={Science},
  volume={378},
  number={6624},
  pages={1092--1097},
  year={2022},
  publisher={American Association for the Advancement of Science}
}

@inproceedings{Fried2023Incoder,
  author       = {Daniel Fried and
                  Armen Aghajanyan and
                  Jessy Lin and
                  Sida Wang and
                  Eric Wallace and
                  Freda Shi and
                  Ruiqi Zhong and
                  Scott Yih and
                  Luke Zettlemoyer and
                  Mike Lewis},
  title        = {InCoder: {A} Generative Model for Code Infilling and Synthesis},
  booktitle    = {{ICLR}},
  publisher    = {OpenReview.net},
  year         = {2023}
}

@article{touvron2023llama,
  title={Llama: Open and efficient foundation language models},
  author={Touvron, Hugo and Lavril, Thibaut and Izacard, Gautier and Martinet, Xavier and Lachaux, Marie-Anne and Lacroix, Timoth{\'e}e and Rozi{\`e}re, Baptiste and Goyal, Naman and Hambro, Eric and Azhar, Faisal and others},
  journal={arXiv preprint arXiv:2302.13971},
  year={2023}
}

@article{roziere2023codellama,
  title={Code llama: Open foundation models for code},
  author={Roziere, Baptiste and Gehring, Jonas and Gloeckle, Fabian and Sootla, Sten and Gat, Itai and Tan, Xiaoqing Ellen and Adi, Yossi and Liu, Jingyu and Remez, Tal and Rapin, J{\'e}r{\'e}my and others},
  journal={arXiv preprint arXiv:2308.12950},
  year={2023}
}

@inproceedings{Nijkamp2023codegen,
  author       = {Erik Nijkamp and
                  Bo Pang and
                  Hiroaki Hayashi and
                  Lifu Tu and
                  Huan Wang and
                  Yingbo Zhou and
                  Silvio Savarese and
                  Caiming Xiong},
  title        = {CodeGen: An Open Large Language Model for Code with Multi-Turn Program
                  Synthesis},
  booktitle    = {{ICLR}},
  publisher    = {OpenReview.net},
  year         = {2023}
}

@article{nijkamp2023codegen2,
  title={Codegen2: Lessons for training llms on programming and natural languages},
  author={Nijkamp, Erik and Hayashi, Hiroaki and Xiong, Caiming and Savarese, Silvio and Zhou, Yingbo},
  journal={arXiv preprint arXiv:2305.02309},
  year={2023}
}

@article{li2023starcoder,
  title={StarCoder: may the source be with you!},
  author={Li, Raymond and Allal, Loubna Ben and Zi, Yangtian and Muennighoff, Niklas and Kocetkov, Denis and Mou, Chenghao and Marone, Marc and Akiki, Christopher and Li, Jia and Chim, Jenny and others},
  journal={arXiv preprint arXiv:2305.06161},
  year={2023}
}

@misc{CodeWhisperer,
author={Amazon},
title={Amazon CodeWhisperer},
howpublished="\url{https://aws.amazon.com/codewhisperer/}",
year={2023},
}

@inproceedings{nguyen2022copilotempirical,
  title={An empirical evaluation of GitHub copilot's code suggestions},
  author={Nguyen, Nhan and Nadi, Sarah},
  booktitle={Proceedings of the 19th International Conference on Mining Software Repositories},
  pages={1--5},
  year={2022}
}

@article{liu2023refining,
  title={Refining ChatGPT-Generated Code: Characterizing and Mitigating Code Quality Issues},
  author={Liu, Yue and Le-Cong, Thanh and Widyasari, Ratnadira and Tantithamthavorn, Chakkrit and Li, Li and Le, Xuan-Bach D and Lo, David},
  journal={arXiv preprint arXiv:2307.12596},
  year={2023}
}

@article{liu2023no,
  title={No Need to Lift a Finger Anymore? Assessing the Quality of Code Generation by ChatGPT},
  author={Liu, Zhijie and Tang, Yutian and Luo, Xiapu and Zhou, Yuming and Zhang, Liang Feng},
  journal={arXiv preprint arXiv:2308.04838},
  year={2023}
}

@article{yeticstiren2023evaluatingquality,
  title={Evaluating the Code Quality of AI-Assisted Code Generation Tools: An Empirical Study on GitHub Copilot, Amazon CodeWhisperer, and ChatGPT},
  author={Yeti{\c{s}}tiren, Burak and {\"O}zsoy, I{\c{s}}{\i}k and Ayerdem, Miray and T{\"u}z{\"u}n, Eray},
  journal={arXiv preprint arXiv:2304.10778},
  year={2023}
}

@article{jesse2023stupidbugs,
  title={Large Language Models and Simple, Stupid Bugs},
  author={Jesse, Kevin and Ahmed, Toufique and Devanbu, Premkumar T and Morgan, Emily},
  journal={arXiv preprint arXiv:2303.11455},
  year={2023}
}

@article{austin2021MBPP,
  title={Program synthesis with large language models},
  author={Austin, Jacob and Odena, Augustus and Nye, Maxwell and Bosma, Maarten and Michalewski, Henryk and Dohan, David and Jiang, Ellen and Cai, Carrie and Terry, Michael and Le, Quoc and others},
  journal={arXiv preprint arXiv:2108.07732},
  year={2021}
}

@inproceedings{lai2023ds1000,
  title={DS-1000: A natural and reliable benchmark for data science code generation},
  author={Lai, Yuhang and Li, Chengxi and Wang, Yiming and Zhang, Tianyi and Zhong, Ruiqi and Zettlemoyer, Luke and Yih, Wen-tau and Fried, Daniel and Wang, Sida and Yu, Tao},
  booktitle={International Conference on Machine Learning},
  pages={18319--18345},
  year={2023},
  organization={PMLR}
}

@article{hendrycks2021apps,
  title={Measuring coding challenge competence with apps},
  author={Hendrycks, Dan and Basart, Steven and Kadavath, Saurav and Mazeika, Mantas and Arora, Akul and Guo, Ethan and Burns, Collin and Puranik, Samir and He, Horace and Song, Dawn and others},
  journal={arXiv preprint arXiv:2105.09938},
  year={2021}
}

@inproceedings{karampatsis2020often,
  title={How often do single-statement bugs occur? the manysstubs4j dataset},
  author={Karampatsis, Rafael-Michael and Sutton, Charles},
  booktitle={Proceedings of the 17th International Conference on Mining Software Repositories},
  pages={573--577},
  year={2020}
}

@misc{chatgpt,
author={OpenAI},
title={ChatGPT: Optimizing Language Models for Dialogue},
howpublished="\url{https://openai.com/blog/chatgpt}",
year={2022},
}

@misc{chatgpt-RL,
author={OpenAI},
title={Learning to Reason with LLMs},
howpublished="\url{https://openai.com/index/learning-to-reason-with-llms/}",
year={2024},
}

@misc{euronews,
author={Euronews},
title={Microsoft attracting users to its code-writing, generative AI software},
howpublished="\url{https://www.euronews.com/next/2023/01/25/microsoft-results-ai}",
year={2023},
}

@misc{Kalliamvakou,
author={Eirini Kalliamvakou},
title={Research: quantifying GitHub Copilot’s impact on developer productivity and happiness},
howpublished="\url{https://github.blog/2022-09-07-research-quantifying-github-copilots-impact-on-developer-productivity-and-happiness/}",
year={2022},
}

@article{maynez2020faithfulness,
  title={On faithfulness and factuality in abstractive summarization},
  author={Maynez, Joshua and Narayan, Shashi and Bohnet, Bernd and McDonald, Ryan},
  journal={arXiv preprint arXiv:2005.00661},
  year={2020}
}

@article{huang2020challenges,
  title={Challenges in building intelligent open-domain dialog systems},
  author={Huang, Minlie and Zhu, Xiaoyan and Gao, Jianfeng},
  journal={ACM Transactions on Information Systems (TOIS)},
  volume={38},
  number={3},
  pages={1--32},
  year={2020},
  publisher={ACM New York, NY, USA}
}

@inproceedings{li2021addressing,
  title={Addressing semantic drift in generative question answering with auxiliary extraction},
  author={Li, Chenliang and Bi, Bin and Yan, Ming and Wang, Wei and Huang, Songfang},
  booktitle={Proceedings of the 59th Annual Meeting of the Association for Computational Linguistics and the 11th International Joint Conference on Natural Language Processing (Volume 2: Short Papers)},
  pages={942--947},
  year={2021}
}

@inproceedings{vaithilingam2022expectation,
  title={Expectation vs. experience: Evaluating the usability of code generation tools powered by large language models},
  author={Vaithilingam, Priyan and Zhang, Tianyi and Glassman, Elena L},
  booktitle={Chi conference on human factors in computing systems extended abstracts},
  pages={1--7},
  year={2022}
}

@inproceedings{Cruzes2011thematic,
  author       = {Daniela S. Cruzes and
                  Tore Dyb{\aa}},
  title        = {Recommended Steps for Thematic Synthesis in Software Engineering},
  booktitle    = {{ESEM}},
  pages        = {275--284},
  publisher    = {{IEEE} Computer Society},
  year         = {2011}
}

@article{guo2024deepseek,
  title={DeepSeek-Coder: When the Large Language Model Meets Programming--The Rise of Code Intelligence},
  author={Guo, Daya and Zhu, Qihao and Yang, Dejian and Xie, Zhenda and Dong, Kai and Zhang, Wentao and Chen, Guanting and Bi, Xiao and Wu, Y and Li, YK and others},
  journal={arXiv preprint arXiv:2401.14196},
  year={2024}
}

@article{tambon2024bugs,
  title={Bugs in Large Language Models Generated Code},
  author={Tambon, Florian and Dakhel, Arghavan Moradi and Nikanjam, Amin and Khomh, Foutse and Desmarais, Michel C and Antoniol, Giuliano},
  journal={arXiv preprint arXiv:2403.08937},
  year={2024}
}

@article{roziere2020unsupervised,
  title={Unsupervised translation of programming languages},
  author={Roziere, Baptiste and Lachaux, Marie-Anne and Chanussot, Lowik and Lample, Guillaume},
  journal={Advances in neural information processing systems},
  volume={33},
  pages={20601--20611},
  year={2020}
}

@article{yuan2023no,
  title={No more manual tests? evaluating and improving chatgpt for unit test generation},
  author={Yuan, Zhiqiang and Lou, Yiling and Liu, Mingwei and Ding, Shiji and Wang, Kaixin and Chen, Yixuan and Peng, Xin},
  journal={arXiv preprint arXiv:2305.04207},
  year={2023}
}

@inproceedings{xia2023automated,
  title={Automated program repair in the era of large pre-trained language models},
  author={Xia, Chunqiu Steven and Wei, Yuxiang and Zhang, Lingming},
  booktitle={2023 IEEE/ACM 45th International Conference on Software Engineering (ICSE)},
  pages={1482--1494},
  year={2023},
  organization={IEEE}
}

@inproceedings{yu2024codereval,
  title={Codereval: A benchmark of pragmatic code generation with generative pre-trained models},
  author={Yu, Hao and Shen, Bo and Ran, Dezhi and Zhang, Jiaxin and Zhang, Qi and Ma, Yuchi and Liang, Guangtai and Li, Ying and Wang, Qianxiang and Xie, Tao},
  booktitle={Proceedings of the 46th IEEE/ACM International Conference on Software Engineering},
  pages={1--12},
  year={2024}
}

@article{huang2023survey,
  title={A survey on hallucination in large language models: Principles, taxonomy, challenges, and open questions},
  author={Huang, Lei and Yu, Weijiang and Ma, Weitao and Zhong, Weihong and Feng, Zhangyin and Wang, Haotian and Chen, Qianglong and Peng, Weihua and Feng, Xiaocheng and Qin, Bing and others},
  journal={arXiv preprint arXiv:2311.05232},
  year={2023}
}

@article{dou2024s,
  title={What's Wrong with Your Code Generated by Large Language Models? An Extensive Study},
  author={Dou, Shihan and Jia, Haoxiang and Wu, Shenxi and Zheng, Huiyuan and Zhou, Weikang and Wu, Muling and Chai, Mingxu and Fan, Jessica and Huang, Caishuang and Tao, Yunbo and others},
  journal={arXiv preprint arXiv:2407.06153},
  year={2024}
}

@article{gupta2020synthesize,
  title={Synthesize, execute and debug: Learning to repair for neural program synthesis},
  author={Gupta, Kavi and Christensen, Peter Ebert and Chen, Xinyun and Song, Dawn},
  journal={Advances in Neural Information Processing Systems},
  volume={33},
  pages={17685--17695},
  year={2020}
}

@article{zhang2023self,
  title={Self-edit: Fault-aware code editor for code generation},
  author={Zhang, Kechi and Li, Zhuo and Li, Jia and Li, Ge and Jin, Zhi},
  journal={arXiv preprint arXiv:2305.04087},
  year={2023}
}

@article{madaan2024self-refine,
  title={Self-refine: Iterative refinement with self-feedback},
  author={Madaan, Aman and Tandon, Niket and Gupta, Prakhar and Hallinan, Skyler and Gao, Luyu and Wiegreffe, Sarah and Alon, Uri and Dziri, Nouha and Prabhumoye, Shrimai and Yang, Yiming and others},
  journal={Advances in Neural Information Processing Systems},
  volume={36},
  year={2024}
}

@article{wei2022chain,
  title={Chain-of-thought prompting elicits reasoning in large language models},
  author={Wei, Jason and Wang, Xuezhi and Schuurmans, Dale and Bosma, Maarten and Xia, Fei and Chi, Ed and Le, Quoc V and Zhou, Denny and others},
  journal={Advances in neural information processing systems},
  volume={35},
  pages={24824--24837},
  year={2022}
}

@article{roit2023factually,
  title={Factually consistent summarization via reinforcement learning with textual entailment feedback},
  author={Roit, Paul and Ferret, Johan and Shani, Lior and Aharoni, Roee and Cideron, Geoffrey and Dadashi, Robert and Geist, Matthieu and Girgin, Sertan and Hussenot, L{\'e}onard and Keller, Orgad and others},
  journal={arXiv preprint arXiv:2306.00186},
  year={2023}
}

@article{wu2024fine,
  title={Fine-grained human feedback gives better rewards for language model training},
  author={Wu, Zeqiu and Hu, Yushi and Shi, Weijia and Dziri, Nouha and Suhr, Alane and Ammanabrolu, Prithviraj and Smith, Noah A and Ostendorf, Mari and Hajishirzi, Hannaneh},
  journal={Advances in Neural Information Processing Systems},
  volume={36},
  year={2024}
}

@article{zhang2023repocoder,
  title={Repocoder: Repository-level code completion through iterative retrieval and generation},
  author={Zhang, Fengji and Chen, Bei and Zhang, Yue and Keung, Jacky and Liu, Jin and Zan, Daoguang and Mao, Yi and Lou, Jian-Guang and Chen, Weizhu},
  journal={arXiv preprint arXiv:2303.12570},
  year={2023}
}

@article{zhao2024retrieval,
  title={Retrieval-augmented generation for ai-generated content: A survey},
  author={Zhao, Penghao and Zhang, Hailin and Yu, Qinhan and Wang, Zhengren and Geng, Yunteng and Fu, Fangcheng and Yang, Ling and Zhang, Wentao and Cui, Bin},
  journal={arXiv preprint arXiv:2402.19473},
  year={2024}
}

@article{peng2023check,
  title={Check your facts and try again: Improving large language models with external knowledge and automated feedback},
  author={Peng, Baolin and Galley, Michel and He, Pengcheng and Cheng, Hao and Xie, Yujia and Hu, Yu and Huang, Qiuyuan and Liden, Lars and Yu, Zhou and Chen, Weizhu and others},
  journal={arXiv preprint arXiv:2302.12813},
  year={2023}
}

@article{chern2023factool,
  title={FacTool: Factuality Detection in Generative AI--A Tool Augmented Framework for Multi-Task and Multi-Domain Scenarios},
  author={Chern, I and Chern, Steffi and Chen, Shiqi and Yuan, Weizhe and Feng, Kehua and Zhou, Chunting and He, Junxian and Neubig, Graham and Liu, Pengfei and others},
  journal={arXiv preprint arXiv:2307.13528},
  year={2023}
}

@inproceedings{DevEval,
  author       = {Jia Li and
                  Ge Li and
                  Yunfei Zhao and
                  Yongmin Li and
                  Huanyu Liu and
                  Hao Zhu and
                  Lecheng Wang and
                  Kaibo Liu and
                  Zheng Fang and
                  Lanshen Wang and
                  Jiazheng Ding and
                  Xuanming Zhang and
                  Yuqi Zhu and
                  Yihong Dong and
                  Zhi Jin and
                  Binhua Li and
                  Fei Huang and
                  Yongbin Li and
                  Bin Gu and
                  Mengfei Yang},
  title        = {DevEval: {A} Manually-Annotated Code Generation Benchmark Aligned
                  with Real-World Code Repositories},
  booktitle    = {{ACL} (Findings)},
  pages        = {3603--3614},
  publisher    = {Association for Computational Linguistics},
  year         = {2024}
}

@inproceedings{Dataflow-RAG,
  author       = {Wei Cheng and
                  Yuhan Wu and
                  Wei Hu},
  title        = {Dataflow-Guided Retrieval Augmentation for Repository-Level Code Completion},
  booktitle    = {{ACL} {(1)}},
  pages        = {7957--7977},
  publisher    = {Association for Computational Linguistics},
  year         = {2024}
}

@article{graphCoder,
  author       = {Wei Liu and
                  Ailun Yu and
                  Daoguang Zan and
                  Bo Shen and
                  Wei Zhang and
                  Haiyan Zhao and
                  Zhi Jin and
                  Qianxiang Wang},
  title        = {GraphCoder: Enhancing Repository-Level Code Completion via Code Context
                  Graph-based Retrieval and Language Model},
  journal      = {CoRR},
  volume       = {abs/2406.07003},
  year         = {2024}
}

@article{Cycle,
  author       = {Yangruibo Ding and
                  Marcus J. Min and
                  Gail E. Kaiser and
                  Baishakhi Ray},
  title        = {{CYCLE:} Learning to Self-Refine the Code Generation},
  journal      = {Proc. {ACM} Program. Lang.},
  volume       = {8},
  number       = {{OOPSLA1}},
  pages        = {392--418},
  year         = {2024}
}

@inproceedings{cov,
  author       = {Shehzaad Dhuliawala and
                  Mojtaba Komeili and
                  Jing Xu and
                  Roberta Raileanu and
                  Xian Li and
                  Asli Celikyilmaz and
                  Jason Weston},
  title        = {Chain-of-Verification Reduces Hallucination in Large Language Models},
  booktitle    = {{ACL} (Findings)},
  pages        = {3563--3578},
  publisher    = {Association for Computational Linguistics},
  year         = {2024}
}

@article{RLCoder,
  author       = {Yanlin Wang and
                  Daya Guo and
                  Jiachi Chen and
                  Ruikai Zhang and
                  Yuchi Ma and
                  Zibin Zheng},
  title        = {RLCoder: Reinforcement Learning for Repository-Level Code Completion},
  journal      = {CoRR},
  volume       = {abs/2407.19487},
  year         = {2024}
}

@inproceedings{CoderReviewer,
  author       = {Zhiyu Li and
                  Shuai Lu and
                  Daya Guo and
                  Nan Duan and
                  Shailesh Jannu and
                  Grant Jenks and
                  Deep Majumder and
                  Jared Green and
                  Alexey Svyatkovskiy and
                  Shengyu Fu and
                  Neel Sundaresan},
  title        = {Automating code review activities by large-scale pre-training},
  booktitle    = {{ESEC/SIGSOFT} {FSE}},
  pages        = {1035--1047},
  publisher    = {{ACM}},
  year         = {2022}
}

@inproceedings{Llama-reviewer,
  author       = {Junyi Lu and
                  Lei Yu and
                  Xiaojia Li and
                  Li Yang and
                  Chun Zuo},
  title        = {LLaMA-Reviewer: Advancing Code Review Automation with Large Language
                  Models through Parameter-Efficient Fine-Tuning},
  booktitle    = {{ISSRE}},
  pages        = {647--658},
  publisher    = {{IEEE}},
  year         = {2023}
}

@article{viera2005understanding,
  title={Understanding interobserver agreement: the kappa statistic},
  author={Viera, Anthony J and Garrett, Joanne M and others},
  journal={Fam med},
  volume={37},
  number={5},
  pages={360--363},
  year={2005}
}

@article{guo2025deepseek,
  title={Deepseek-r1: Incentivizing reasoning capability in llms via reinforcement learning},
  author={Guo, Daya and Yang, Dejian and Zhang, Haowei and Song, Junxiao and Zhang, Ruoyu and Xu, Runxin and Zhu, Qihao and Ma, Shirong and Wang, Peiyi and Bi, Xiao and others},
  journal={arXiv preprint arXiv:2501.12948},
  year={2025}
}

@article{fitzgerald2015important,
  title={Important text characteristics for early-grades text complexity.},
  author={Fitzgerald, Jill and Elmore, Jeff and Koons, Heather and Hiebert, Elfrieda H and Bowen, Kimberly and Sanford-Moore, Eleanor E and Stenner, A Jackson},
  journal={Journal of Educational Psychology},
  volume={107},
  number={1},
  pages={4},
  year={2015},
  publisher={American Psychological Association}
}

@article{fitzgerald2016examining,
  title={Examining text complexity in the early grades},
  author={Fitzgerald, Jill and Elmore, Jeff and Hiebert, Elfrieda H and Koons, Heather H and Bowen, Kimberly and Sanford-Moore, Eleanor E and Stenner, A Jackson},
  journal={Phi Delta Kappan},
  volume={97},
  number={8},
  pages={60--65},
  year={2016},
  publisher={SAGE Publications Sage CA: Los Angeles, CA}
}

@inproceedings{rasnayaka2024empirical,
  title={An empirical study on usage and perceptions of llms in a software engineering project},
  author={Rasnayaka, Sanka and Wang, Guanlin and Shariffdeen, Ridwan and Iyer, Ganesh Neelakanta},
  booktitle={Proceedings of the 1st International Workshop on Large Language Models for Code},
  pages={111--118},
  year={2024}
}

@article{tabarsi2025llms,
  title={LLMs' Reshaping of People, Processes, Products, and Society in Software Development: A Comprehensive Exploration with Early Adopters},
  author={Tabarsi, Benyamin and Reichert, Heidi and Limke, Ally and Kuttal, Sandeep and Barnes, Tiffany},
  journal={arXiv preprint arXiv:2503.05012},
  year={2025}
}

@inproceedings{zhong2024can,
  title={Can llm replace stack overflow? a study on robustness and reliability of large language model code generation},
  author={Zhong, Li and Wang, Zilong},
  booktitle={Proceedings of the AAAI Conference on Artificial Intelligence},
  volume={38},
  number={19},
  pages={21841--21849},
  year={2024}
}

@inproceedings{sharma2024llms,
  title={LLMs for Code: The Potential, Prospects, and Problems},
  author={Sharma, Tushar},
  booktitle={2024 IEEE 21st International Conference on Software Architecture Companion (ICSA-C)},
  pages={373--374},
  year={2024},
  organization={IEEE}
}

@inproceedings{lee2022deduplicating,
  title={Deduplicating Training Data Makes Language Models Better},
  author={Lee, Katherine and Ippolito, Daphne and Nystrom, Andrew and Zhang, Chiyuan and Eck, Douglas and Callison-Burch, Chris and Carlini, Nicholas},
  booktitle={Proceedings of the 60th Annual Meeting of the Association for Computational Linguistics (Volume 1: Long Papers)},
  pages={8424--8445},
  year={2022}
}

@article{onoe2022entity,
  title={Entity cloze by date: What LMs know about unseen entities},
  author={Onoe, Yasumasa and Zhang, Michael JQ and Choi, Eunsol and Durrett, Greg},
  journal={arXiv preprint arXiv:2205.02832},
  year={2022}
}

@inproceedings{zhang2024language,
  title={How Language Model Hallucinations Can Snowball},
  author={Zhang, Muru and Press, Ofir and Merrill, William and Liu, Alisa and Smith, Noah A},
  booktitle={International Conference on Machine Learning},
  pages={59670--59684},
  year={2024},
  organization={PMLR}
}

@inproceedings{holtzman2020curious,
  author       = {Ari Holtzman and
                  Jan Buys and
                  Li Du and
                  Maxwell Forbes and
                  Yejin Choi},
  title        = {The Curious Case of Neural Text Degeneration},
  booktitle    = {{ICLR}},
  publisher    = {OpenReview.net},
  year         = {2020}
}

@article{shim2024cot,
  title={CoT Harms Performance of Rather Smaller Language Models},
  author={Shim, Jihoo and Ho, Shin Dong and Kim, Jeongwon},
  year={2024}
}

@article{lewis2020retrieval,
  title={Retrieval-augmented generation for knowledge-intensive nlp tasks},
  author={Lewis, Patrick and Perez, Ethan and Piktus, Aleksandra and Petroni, Fabio and Karpukhin, Vladimir and Goyal, Naman and K{\"u}ttler, Heinrich and Lewis, Mike and Yih, Wen-tau and Rockt{\"a}schel, Tim and others},
  journal={Advances in neural information processing systems},
  volume={33},
  pages={9459--9474},
  year={2020}
}

@article{zhang2024llm,
  title={Llm hallucinations in practical code generation: Phenomena, mechanism, and mitigation},
  author={Zhang, Ziyao and Wang, Yanlin and Wang, Chong and Chen, Jiachi and Zheng, Zibin},
  journal={arXiv preprint arXiv:2409.20550},
  year={2024}
}

@article{spracklen2024we,
  title={We have a package for you! A comprehensive analysis of package hallucinations by code generating llms},
  author={Spracklen, Joseph and Wijewickrama, Raveen and Sakib, AHM and Maiti, Anindya and Viswanath, Bimal and Jadliwala, Murtuza},
  journal={arXiv preprint arXiv:2406.10279},
  year={2024}
}

@article{agarwal2024codemirage,
  title={Codemirage: Hallucinations in code generated by large language models},
  author={Agarwal, Vibhor and Pei, Yulong and Alamir, Salwa and Liu, Xiaomo},
  journal={arXiv preprint arXiv:2408.08333},
  year={2024}
}

@article{jiang2024collu,
  title={Collu-Bench: A Benchmark for Predicting Language Model Hallucinations in Code},
  author={Jiang, Nan and Li, Qi and Tan, Lin and Zhang, Tianyi},
  journal={arXiv preprint arXiv:2410.09997},
  year={2024}
}

@article{rahman2024code,
  title={Code Hallucination},
  author={Rahman, Mirza Masfiqur and Kundu, Ashish},
  journal={arXiv preprint arXiv:2407.04831},
  year={2024}
}

@article{tian2024codehalu,
  title={Codehalu: Code hallucinations in llms driven by execution-based verification},
  author={Tian, Yuchen and Yan, Weixiang and Yang, Qian and Chen, Qian and Wang, Wen and Luo, Ziyang and Ma, Lei},
  journal={arXiv e-prints},
  pages={arXiv--2405},
  year={2024}
}

@article{mesbahreducing,
  title={Reducing Hallucinations: Harnessing Contextual Analysis for AI Code Generation},
  author={Mesbah, Ali}
}

@article{liu2023your,
  title={Is your code generated by chatgpt really correct? rigorous evaluation of large language models for code generation},
  author={Liu, Jiawei and Xia, Chunqiu Steven and Wang, Yuyao and Zhang, Lingming},
  journal={Advances in Neural Information Processing Systems},
  volume={36},
  pages={21558--21572},
  year={2023}
}

@article{dhuliawala2023chain,
  title={Chain-of-verification reduces hallucination in large language models},
  author={Dhuliawala, Shehzaad and Komeili, Mojtaba and Xu, Jing and Raileanu, Roberta and Li, Xian and Celikyilmaz, Asli and Weston, Jason},
  journal={arXiv preprint arXiv:2309.11495},
  year={2023}
}

@article{li2023inference,
  title={Inference-time intervention: Eliciting truthful answers from a language model},
  author={Li, Kenneth and Patel, Oam and Vi{\'e}gas, Fernanda and Pfister, Hanspeter and Wattenberg, Martin},
  journal={Advances in Neural Information Processing Systems},
  volume={36},
  pages={41451--41530},
  year={2023}
}

@inproceedings{xie2023adaptive,
  title={Adaptive chameleon or stubborn sloth: Revealing the behavior of large language models in knowledge conflicts},
  author={Xie, Jian and Zhang, Kai and Chen, Jiangjie and Lou, Renze and Su, Yu},
  booktitle={The Twelfth International Conference on Learning Representations},
  year={2023}
}

@article{weijia2023replug,
  title={REPLUG: Retrieval-augmented black-box language models},
  author={Weijia, Shi and Sewon, Min and Michihiro, Yasunaga and Minjoon, Seo and Rich, James and Mike, Lewis and Wen-tau, Yih},
  journal={arXiv preprint ArXiv:2301.12652},
  year={2023}
}

@inproceedings{du2023improving,
  title={Improving factuality and reasoning in language models through multiagent debate},
  author={Du, Yilun and Li, Shuang and Torralba, Antonio and Tenenbaum, Joshua B and Mordatch, Igor},
  booktitle={Forty-first International Conference on Machine Learning},
  year={2023}
}

@article{rebuffel2022controlling,
  title={Controlling hallucinations at word level in data-to-text generation},
  author={Rebuffel, Cl{\'e}ment and Roberti, Marco and Soulier, Laure and Scoutheeten, Geoffrey and Cancelliere, Rossella and Gallinari, Patrick},
  journal={Data Mining and Knowledge Discovery},
  pages={1--37},
  year={2022},
  publisher={Springer}
}

@article{li2023batgpt,
  title={Batgpt: A bidirectional autoregessive talker from generative pre-trained transformer},
  author={Li, Zuchao and Zhang, Shitou and Zhao, Hai and Yang, Yifei and Yang, Dongjie},
  journal={arXiv preprint arXiv:2307.00360},
  year={2023}
}

@inproceedings{chendycodeeval,
  title={DyCodeEval: Dynamic Benchmarking of Reasoning Capabilities in Code Large Language Models Under Data Contamination},
  author={Chen, Simin and Pusarla, Pranav and Ray, Baishakhi},
  booktitle={Forty-second International Conference on Machine Learning}
}

@article{chen2024ppm,
  title={Ppm: Automated generation of diverse programming problems for benchmarking code generation models},
  author={Chen, Simin and Feng, Xiaoning and Han, Xiaohong and Liu, Cong and Yang, Wei},
  journal={Proceedings of the ACM on Software Engineering},
  volume={1},
  number={FSE},
  pages={1194--1215},
  year={2024},
  publisher={ACM New York, NY, USA}
}

@article{jain2024livecodebench,
  title={Livecodebench: Holistic and contamination free evaluation of large language models for code},
  author={Jain, Naman and Han, King and Gu, Alex and Li, Wen-Ding and Yan, Fanjia and Zhang, Tianjun and Wang, Sida and Solar-Lezama, Armando and Sen, Koushik and Stoica, Ion},
  journal={arXiv preprint arXiv:2403.07974},
  year={2024}
}

@article{zhuo2024bigcodebench,
  title={Bigcodebench: Benchmarking code generation with diverse function calls and complex instructions},
  author={Zhuo, Terry Yue and Vu, Minh Chien and Chim, Jenny and Hu, Han and Yu, Wenhao and Widyasari, Ratnadira and Yusuf, Imam Nur Bani and Zhan, Haolan and He, Junda and Paul, Indraneil and others},
  journal={arXiv preprint arXiv:2406.15877},
  year={2024}
}

@article{khan2023xcodeeval,
  title={xcodeeval: A large scale multilingual multitask benchmark for code understanding, generation, translation and retrieval},
  author={Khan, Mohammad Abdullah Matin and Bari, M Saiful and Do, Xuan Long and Wang, Weishi and Parvez, Md Rizwan and Joty, Shafiq},
  journal={arXiv preprint arXiv:2303.03004},
  year={2023}
}

\end{document}